\documentclass{aa}
\usepackage[utf8]{inputenc}

\usepackage{graphicx}
\usepackage{txfonts}
\usepackage{natbib} 
\usepackage{lscape} 
\usepackage{subcaption} 

\newcommand{\mjup}{M$_{\text{J}}$}
\newcommand{\rjup}{R$_{\text{J}}$}

\newcommand{\ms}{$\textrm{ms}^{-1}$}

\begin{document}
\title{Direct imaging of an ultracool substellar companion to the exoplanet host star HD\,4113\,A \thanks{Based on observations collected at the European Organisation for Astronomical Research in the Southern Hemisphere under ESO programmes 097.C-0893(A), 077.C-0293(A), 279.C-5052(A), 081.C-0653(A) and 091.C-0721(B).}}

\author{A. Cheetham\inst{1}
          \and
          D. S{\'e}gransan\inst{1}
          \and
          S. Peretti\inst{1}
          \and
          J.-B. Delisle\inst{1}
          \and
          J. Hagelberg\inst{2,3}
          \and
          J-L. Beuzit\inst{2}
           \and
          T. Forveille\inst{2}
          \and
          M. Marmier\inst{1}
          \and
          S. Udry\inst{1}
          \and
          F. Wildi\inst{1}
}
   \institute{D{\'e}partement d'Astronomie, Universit{\'e} de Gen{\`e}ve, 51 chemin des Maillettes, 1290, Versoix, Switzerland
             \email{anthony.cheetham@unige.ch}
        \and
          Universit{\'e} Grenoble Alpes, CNRS, IPAG, 38000 Grenoble, France
        \and
        Institute for Astronomy, University of Hawai'i, 2680 Woodlawn Drive, Honolulu, HI 96822, USA
             }
\date{}

\abstract{Using high-contrast imaging with the SPHERE instrument at the VLT, we report the first images of a cold brown dwarf companion to the exoplanet host star HD\,4113A. The brown dwarf HD\,4113C is part of a complex dynamical system consisting of a giant planet, stellar host and a known wide M-dwarf companion. Its separation of $535\pm3$\,mas and H-band contrast of $13.35\pm0.10$\,mag correspond to a projected separation of 22\,AU and an isochronal mass estimate of $36\pm5$\,\mjup\, based on COND models. The companion shows strong methane absorption, and through atmospheric model fitting we estimate a surface gravity of $\log g$=5 and an effective temperature of $\sim$500-600\,K. A comparison of its spectrum with observed T dwarfs indicates a late-T spectral type, with a T9 object providing the best match. By combining the observed astrometry from the imaging data with 27 years of radial velocities, we use orbital fitting to constrain its orbital and physical parameters, as well as update those of the planet HD\,4113A\,b, discovered by previous radial velocity measurements. The data suggest a dynamical mass of $66^{+5}_{-4}$\mjup\, and moderate eccentricity of $0.44^{+0.08}_{-0.07}$ for the brown dwarf. This mass estimate appears to conflict with the isochronal estimate and that of objects with similar temperatures, which may be caused by the newly detected object being an unresolved binary brown dwarf system or the presence of an additional object in the system.
Through dynamical simulations we show that the planet may undergo strong Lidov-Kozai cycles, raising the possibility that it formed on a quasi-circular orbit, and gained its currently observed high eccentricity ($e\sim0.9$) through interactions with the brown dwarf. 
Follow-up observations combining radial velocities, direct imaging and Gaia astrometry will be crucial to precisely constrain the dynamical mass of the brown dwarf and allow for in-depth comparison with evolutionary and atmospheric models.

}


\titlerunning{Direct imaging of a substellar companion to HD\,4113\,A}

\maketitle

\section{Introduction}

HD\,4113A is a nearby G5 dwarf that hosts a giant planet with one of the highest known orbital eccentricities (e=0.9). The planetary companion, HD\,4113A\,b, was first detected via radial velocity (RV) measurements by \citet{2008A&A...480L..33T} using the \textsc{CORALIE} spectrograph on the 1.2\,m Swiss telescope at La Silla Observatory, as part of a planet search survey of a well-defined sample of solar-type stars \citep{2000A&A...356..590U}.

One common explanation for the presence of such a high eccentricity is through Lidov-Kozai \citep{kozai_secular_1962,lidov_evolution_1962} interactions with a companion on a wider orbit. The RV measurements presented by \cite{2008A&A...480L..33T} also revealed a long-term trend, indicative of such a companion with a much longer orbital period. These data suggested a minimum mass of 10\,\mjup\, and minimum semi-major axis of 8\,AU. Using this knowledge, we made several unsuccessful attempts to image this companion, first, with the NACO instrument at the VLT, using H-band Spectral Differential Imaging \citep[SDI]{1987PASP...99.1344S} in 2006 and 2007 \citep{montagnier:tel-00714874}, coronagraphy in 2008 and non-coronagraphic L-band ADI in 2013 \citep{hagelberg:unige-48340}. 
A summary of the L-band ADI results is presented in Appendix \ref{appendixA}.
These observations ruled out stellar companions between 0.2-2.5\,arcsec (8-100\,AU), suggesting that the outer companion was a brown dwarf.

In addition, a comoving stellar companion to HD\,4113A was discovered by \citet{2014MNRAS.439.1063M} at a much larger separation (49\,arcsec, $~$2000\,AU) and was identified as an early M dwarf using near-IR photometry. The low mass of HD\,4113B and its extremely long period cannot explain the observed RV drift. The authors also excluded the presence of any additional stellar companions between 190 and 6500\,AU down to the hydrogen-burning mass limit.

While these previous imaging observations significantly narrowed down the parameter space to detect the unseen companion, the most recent generation of extreme adaptive optics instruments such as SPHERE \citep{2008SPIE.7014E..18B} and GPI \citep{Macintosh2008SPIE} are able to obtain higher contrasts at smaller angular separations, opening the search to even lower masses.

In this paper we report the first direct images of the suspected companion HD\,4113C, confirming its presumed nature as a brown dwarf. In addition we extend the time baseline of the radial velocity observations by 6 years. When combined with the imaging data, this allows constraints to be placed on the mass and orbital parameters of the brown dwarf and planetary companion.

The paper is organized as follows: In Section \ref{stellarParams}, we revisit the stellar parameters of HD\,4113A+B, using the most recent available data to better constrain their mass and age. In Section \ref{sec:observations_and_reduction}, we summarize the observations and data analysis procedures for the radial velocity and high contrast imaging data. In Section \ref{sec:results}, we present the results of the imaging data analysis and derived companion properties, an overview of the results from the combined orbital fitting, as well as the results of dynamical simulations showing the interactions between the brown dwarf and giant planet. We conclude in Section \ref{sec:conclusion} with a summary of the results presented and their implications.

\section{Stellar parameters\label{stellarParams}}
Following the recent Gaia Data Release 1 \citep{2016arXiv160904172G}, we are able to refine the stellar parameters of HD\,4113A+B. This data gives an improved parallax of $\pi=24.0\pm0.2$ mas, setting the star at a slightly closer distance to the Sun than previously thought (d= $41.7\pm0.7$ pc\footnote{A systematic error of 0.4312\,mas - corresponding to the astrometric excess noise measured by GAIA -  was quadratically added to the parallax error to absorb a suspected bias linked to the signature of HD\,4113A\,b in the GAIA data.}).

For HD\,4113A, we combined the  Gaia parallaxes with the Hipparcos and 2MASS photometry \citep{2006AJ....131.1163S} to derive the absolute magnitudes listed in Table \ref{tab-mcmc-StellarParams}. $T_\text{eff}$, $\log g$ and [Fe/H] are taken from the spectroscopic analysis done in \citet{2008A&A...480L..33T} while the $v\,sin{i}$ and the $\log{R^{'}_{HK}}$ are derived from the CORALIE-07 spectra following the approach of \citet{santos_2002} and \citet{2011arXiv1107.5325L} and implemented in CORALIE by \cite{Marmier_phd}.
The fundamental parameters of this metal rich star ([Fe/H]$=0.20 \pm 0.04$) were derived using the Geneva stellar evolution models \citep{2012A&A...541A..41M}. The interpolation in the model grid was done through a Bayesian formalism using observational Gaussian priors on $T_\text{eff}$, $M_{V}$, $\log{g}$ and [Fe/H]    \citep{Marmier_phd}. We derived a mass of $M=1.05_{-0.02}^{+0.04} M_{\odot}$ and an age of  $5.0_{-1.7}^{+1.3}$\,Gyr.

While HD\,4113B is likely to have similar metal enrichment as HD\,4113A, \citet{2000A&A...364..217D} showed that the Mass-Luminosity relation for M dwarfs in the near-infrared is only marginally affected by metallicity. For this reason we use the Solar metallicity  model grids of \citet{2015A&A...577A..42B} combined with the 2MASS and Sofi/ESO NTT  near-infrared photometry  \citep{2014MNRAS.439.1063M} to derive its stellar parameters. 
The interpolation in the model grid was done through a Bayesian formalism using observational Gaussian priors on $M_{J}$, $M_{H}$, $M_{K}$ and the $H-K$ color index. Through this approach we obtain a stellar mass of $M_{B}=0.55 M_{\odot}$ and an effective temperature of $T_\text{eff}=3833$\,K, in agreement with a M0-1V spectral type \citep{2013A&A...556A..15R}. All observed and derived stellar parameters are  given in Table \ref{tab-mcmc-StellarParams}.

\begin{table} 
\centering
\caption{Stellar Parameters of HD\,4113 A \& B }  
\label{tab-mcmc-StellarParams}  
\begin{tabular}{lccc}  
\hline    
Param&units&HD\,4113A&HD\,4113B\\
\hline  
SpType & & G5V &M0-1V\\
V      & &7.880 $\pm$ 0.013 &12.70 $\pm$ 0.02\\
J      & &6.672 $\pm$ 0.023 &9.447 $\pm$ 0.024\\
H      & &6.345 $\pm$ 0.021 & 8.74 $\pm$ 0.02\\
K      & &6.289 $\pm$ 0.021 & 8.539 $\pm$ 0.021\\
B-V    & &0.716 $\pm$ 0.003 & - \\
\hline  
$\pi$  & [mas]&\multicolumn{2}{c}{23.986 $\pm$ 0.237 $\pm$ 0.4312(syst)}\\
d      & [pc]&\multicolumn{2}{c}{41.70 $\pm$ 0.86}\\
\hline  
$M_{V}$&     & 4.780 $\pm$ 0.046&9.600 $\pm$ 0.049\\
$M_{J}$&     & 3.572 $\pm$ 0.050&6.347 $\pm$ 0.051\\
$M_{H}$&     & 3.274 $\pm$ 0.045&5.631 $\pm$ 0.045\\
$M_{K}$&     & 3.234 $\pm$ 0.045&5.445 $\pm$ 0.045\\
\hline  
$T_\text{eff}$   & [K]&5688 $\pm$ 26 $\pm$ 50(syst) &3833\\
$[$Fe/H$]$    & [dex]&+0.20 $\pm$ 0.04&- \\
$\log g$ &      &4.40 $\pm$ 0.05&4.76 \\
$\log{R^{'}_{HK}}$& &-5.104 $\pm$ 0.003&-\\
$v\,\sin{i}$& [km/s] &2.324 &-\\
\hline  
Age    & [Gyr]&$5.0_{-1.7}^{+1.3}$&5(fixed)\\
$M_{A}$ &[M$_{\odot}$] &$1.05_{-0.02}^{+0.04}$&0.55\\ 
\hline                  
\end{tabular}
\end{table}

\section{Observations and Data Reduction\label{sec:observations_and_reduction}}

\subsection{Radial velocities}
\begin{figure}
\centering
  \includegraphics[width=0.95\columnwidth]{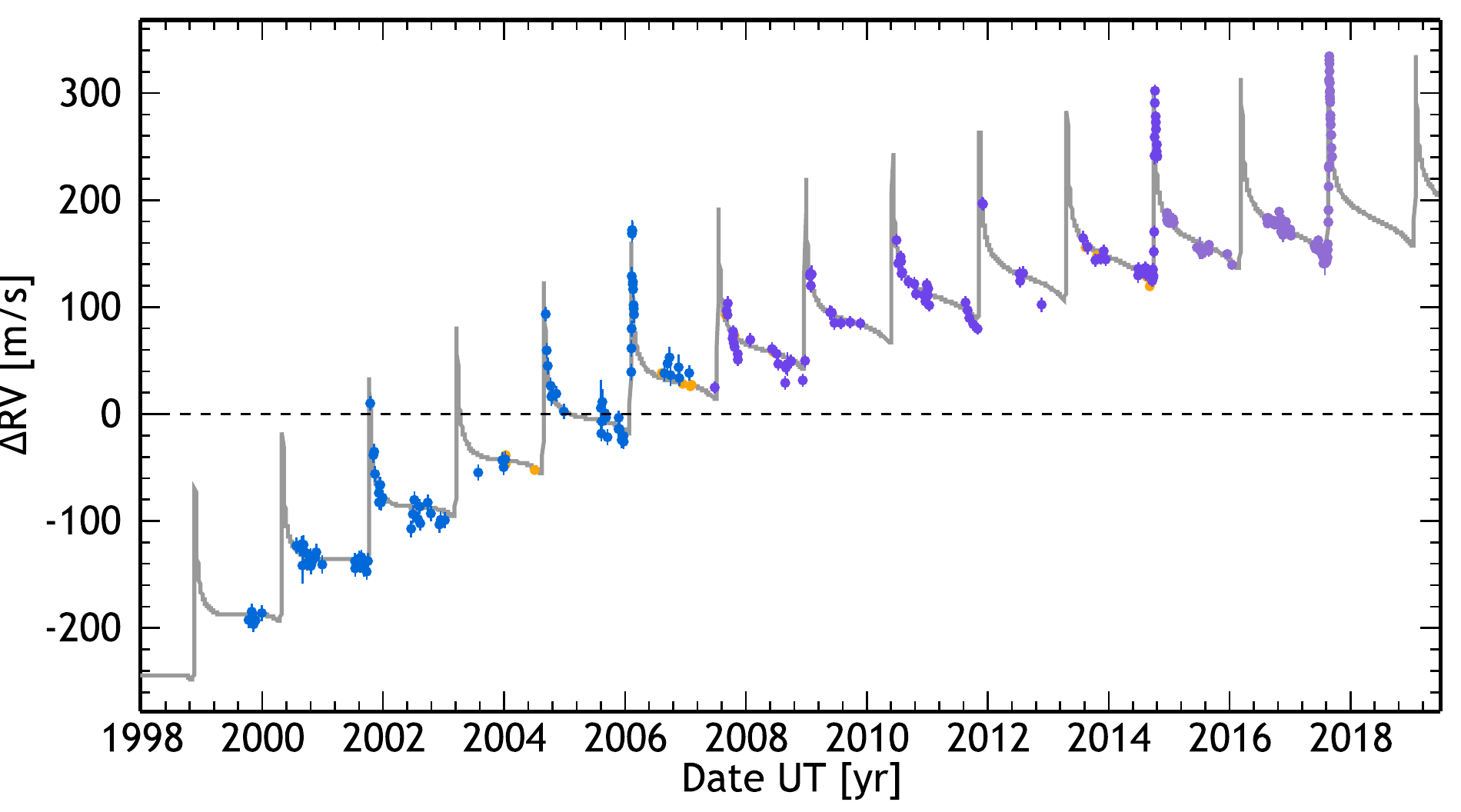}
  \includegraphics[width=0.95\columnwidth]{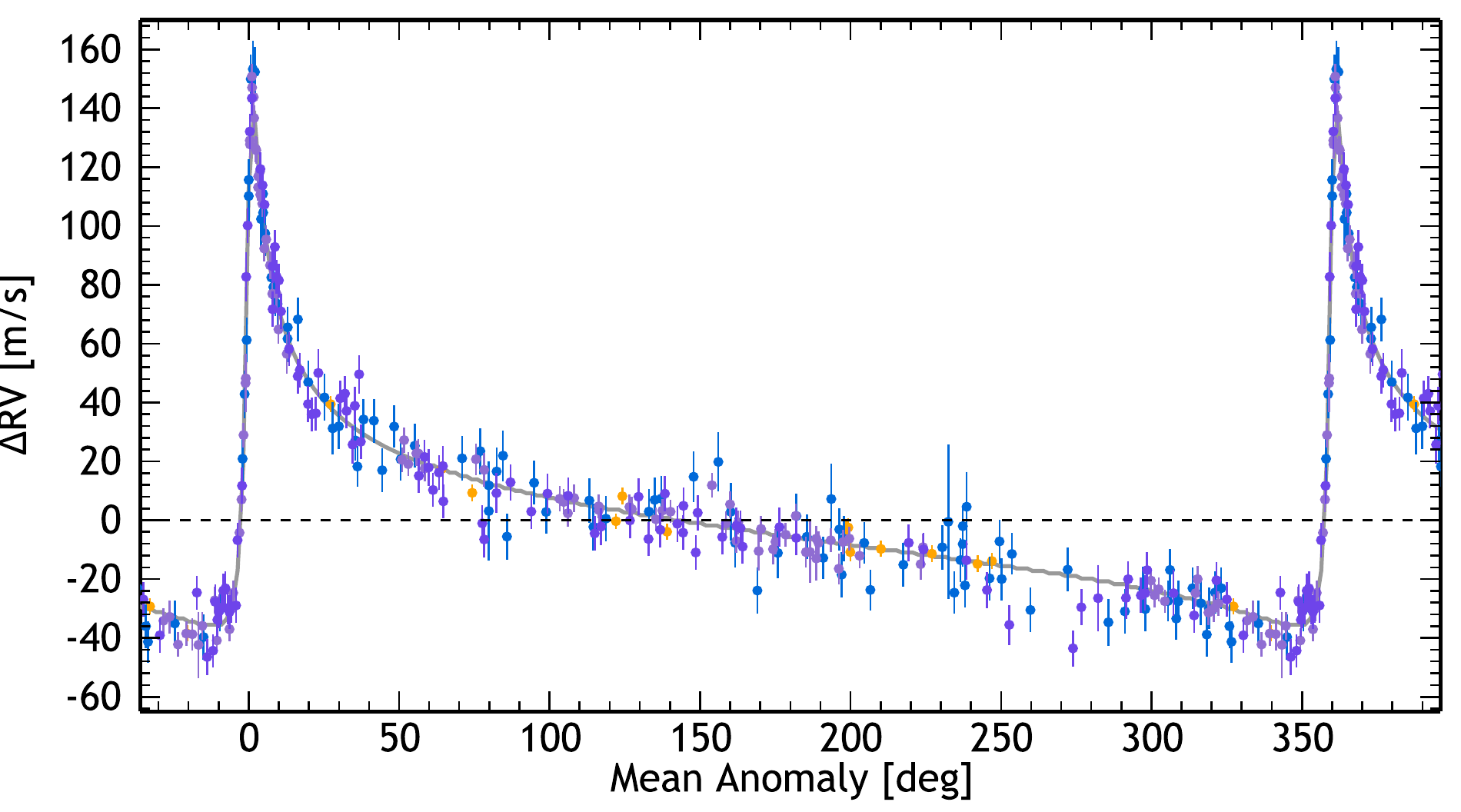}
  \includegraphics[width=0.95\columnwidth]{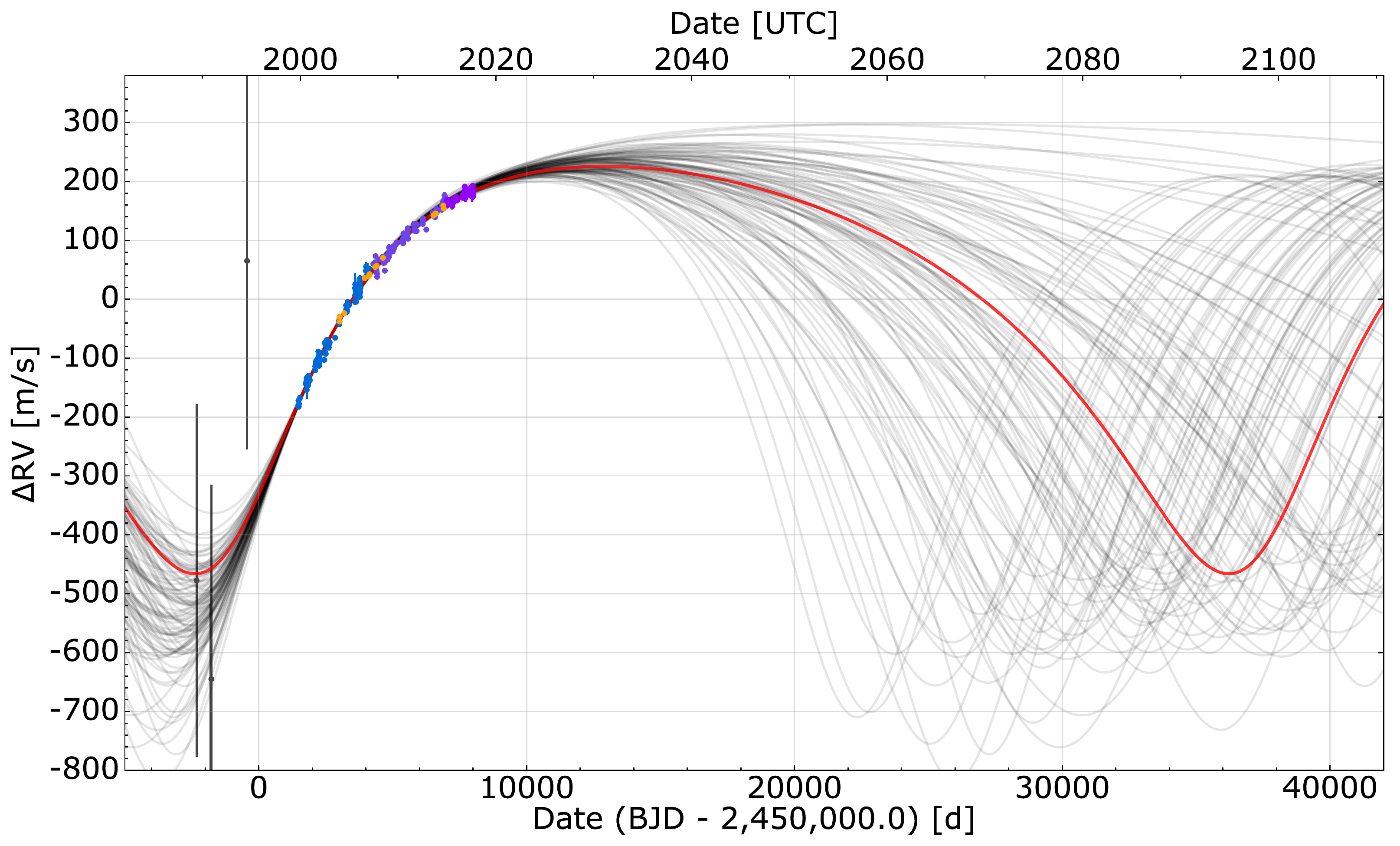}
  \caption{Radial velocity measurements of HD\,4113A taken with \textsc{CORALIE} (blue dots for data taken prior to the 2007 upgrade and purple dots for data taken later) and  with HIRES (orange dots) covering the time period from October 1999 to September 2017. Four CORAVEL measurements from as early as May 1989 are also included (black dots), extending the time baseline to 28 years. The top figure represents the observed CORALIE RV time series with the eccentric orbit model of the giant planet HD\,4113A\,b as well as the long term drift corresponding to HD\,4113\,C. The middle figure shows a phase-folded representation of the RV orbit of HD\,4113A\,b, while the bottom figure shows the predicted and observed RVs for the orbit of HD\,4113C, after subtraction of the signal produced by HD\,4113A\,b. A range of possible orbits drawn from the output of our MCMC procedure are shown. The bold red curve represents the Maximum likelihood solution.}
  \label{fig:rvs}
\end{figure}

Since the discovery of HD\,4113A\,b by \citet{2008A&A...480L..33T}, we have acquired more than 166 additional radial velocity measurements of HD\,4113A, with the goal of better characterizing the long period and more massive companion responsible for the observed radial velocity drift. The 272 measurements obtained between October 1999 and September 2017 are shown in Fig. \ref{fig:rvs} with the corresponding Keplerian model for HD\,4113A\,b and the long term drift.
During that period of time, \textsc{CORALIE} underwent two upgrades that introduced small zero point RV offsets \citep{2010A&A...511A..45S}. To account for these, we consider the data as originating from three different instruments with unknown offsets: \textsc{CORALIE-98}, \textsc{CORALIE-07} and \textsc{CORALIE-14}. These values are then included in the orbital fit as additional parameters.
We also included 16 \textsc{KECK-HIRES} RV measurements \citep{2017AJ....153..208B} covering a 10 year time span from 2004 to 2014.
Four unpublished measurements from the CORAVEL spectrograph taken between May 1989 and July 1994 were also added. While these measurements have large uncertainties, the significant increase in time baseline helps to constrain the possible amplitude and period of the signal introduced by the long period companion responsible for the RV drift\footnote{Due to their large error bars, CORAVEL measurements are only shown on the bottom panel of Fig. \ref{fig:rvs} only.}.

\subsection{2016 SPHERE high-contrast imaging}

HD\,4113A was observed with SPHERE, the extreme adaptive optics system at the VLT \citep{2008SPIE.7014E..18B} on 2016-07-20 (Program 097.C-0893, PI Cheetham). Observations were taken in IRDIFS mode, which allows the Integral Field Spectrograph \citep[IFS;][]{2015A&A...576A.121M} and InfraRed Dual-band Imager and Spectrograph \citep[IRDIS;][]{2008SPIE.7014E..3LD} modules to be used simultaneously. The IFS data cover a range of wavelengths from 0.96-1.34\,$\mu$m with an average spectral resolution of 55.1. The IRDIS data were taken in the dual-band imaging mode \citep{2010MNRAS.407...71V} using the H2 and H3 filters ($\lambda_{H_2}=1.5888\mu$m, $\lambda_{H_3}=1.6671\mu$m), designed such that the H3 filter aligns with a methane absorption band. The conditions during the observations were good, with stable wind and seeing, but with some intermittent clouds that may affect the photometry.

The observing sequence consisted of long exposure images taken with an apodized Lyot coronagraph \citep{2003A&A...397.1161S} with a spot diameter of 185\,mas. To measure the position of the star behind the coronagraph, several exposures were taken with a sinusoidal modulation applied to the deformable mirror (to generate satellite spots around the star) at the beginning and end of the sequence. To estimate the stellar flux and the shape of the Point Spread Function (PSF) during the sequence, several short exposure images were taken with the star moved from behind the coronagraph and using a neutral-density filter with a $\sim$10\% transmission, also at the beginning and end of the sequence. In addition, several long exposure sky frames were taken to estimate the background flux and help identify bad pixels on the detector. The observing sequence included a maximum of 54\,deg of field rotation between the coronagraphic frames. The observing parameters are given in Table \ref{tab:observing_log}.

\begin{table*}
\centering
\caption{Observing Log for the 2 SPHERE datasets. The "flux" columns refer to the short exposure images of HD\,4113A used as a PSF reference and to measure the flux of the primary.}
\begin{tabular}{ccc|cc|ccc|cc}
 & & & \multicolumn{2}{c|}{Coronagraphic sequence} & \multicolumn{3}{c|}{Photometric sequences} & \multicolumn{2}{c}{Sky sequence}\\
Camera & Date & Filter/Prism & DIT & Exposures & DIT &
Exposures & ND filter & DIT & Exposures \\
\hline
IRDIS & 2016-07-20 & H2 + H3 & 64s & 64 & 0.84s & 40 & ND 1.0 & 64 & 2 \\
IFS & 2016-07-20 & Y-J & 64s & 64 & 2s & 20 & ND 1.0 & 64 & 2 \\
IRDIS & 2015-10-08 & K1 + K2 & 32s & 56 & 0.84s & 4 & ND 1.0 & 64 & 2 \\
IFS & 2015-10-08 & Y-H & 32s & 56 & 2s & 4 & ND 1.0 & 64 & 2 \\

\end{tabular}
\label{tab:observing_log}
\end{table*}


The SPHERE Data Reduction and Handling pipeline \citep[v0.18.0]{2008SPIE.7019E..39P} was used to perform the wavelength extraction for the IFS data, turning the full-frame images of the lenslet spectra into image cubes.

The remainder of the data reduction and analysis procedure was completed using the GRAPHIC pipeline \citep{2016MNRAS.455.2178H}, with recent modifications to handle SPHERE data. The data were first sky subtracted, flat fielded, cleaned of bad pixels and corrected for distortion (following \cite{2016arXiv160906681M}). The position of the primary star was measured from the images with the satellite spots, and the average value was used as the assumed position. The measured difference between the values measured at the start and end of the sequence was 2.5\,mas. Coronagraphic images were then shifted to the measured centre position using Fourier transforms.

An implementation of the Principal Component Analysis (PCA) based KLIP PSF subtraction algorithm \citep{2012ApJ...755L..28S} was then run separately for images at each wavelength and for each IRDIS channel. The resulting frames were derotated and median combined to produce a final PSF-subtracted image. The PCA algorithm was applied separately to annular sections of each image. Each annulus had a width of 2 FWHM (calculated for each channel individually). One annulus was centred on the observed position of the companion and the remaining annuli were defined inwards and outwards until reaching the edge of the coronagraph and the edge of the field of view. To minimize companion self-subtraction, we excluded images with small changes in parallactic angle when building a reference library for each image and annulus. Our chosen criteria required that a companion in the centre of each annulus would move by a minimum of 1.25$\times$ the PSF FWHM due to sky rotation.

\begin{figure*}
\centering

  \includegraphics[width=0.9\textwidth]{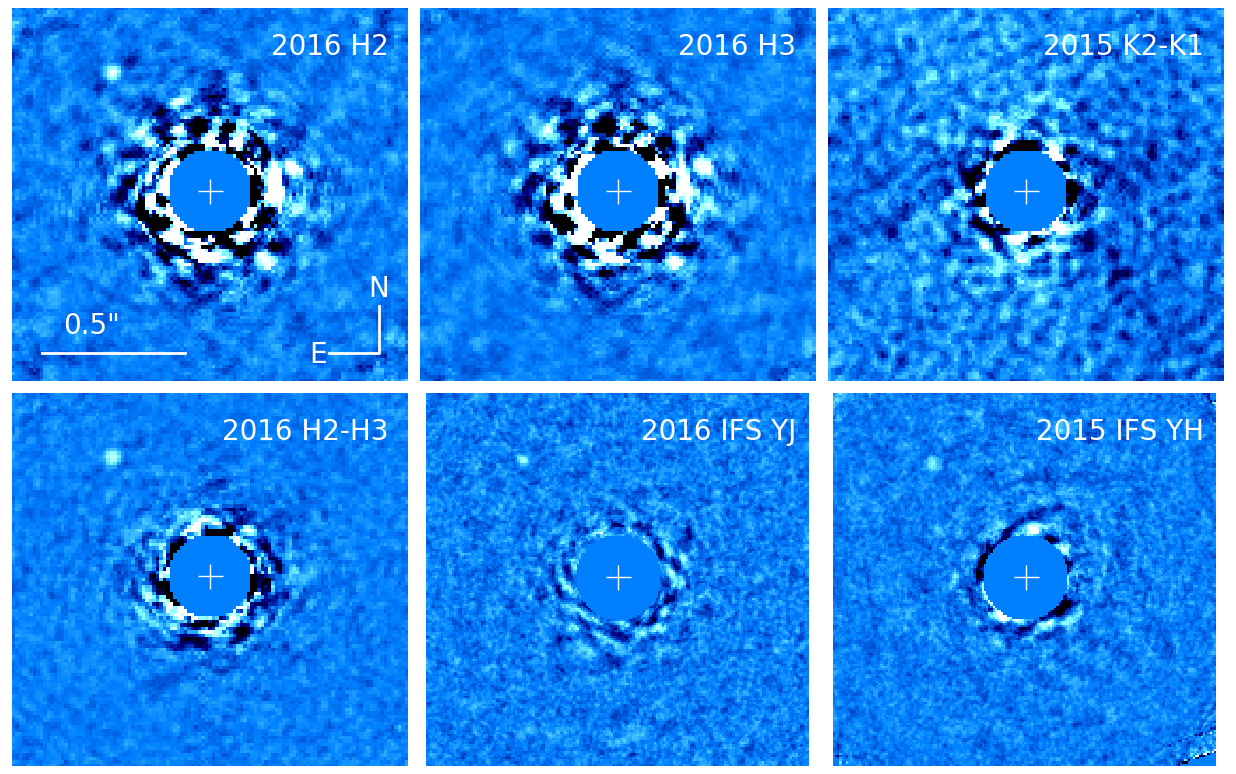}

  \caption{The PSF-subtracted images from each dataset and epoch. For the 2016 dataset, the IRDIS H2, H3 and SDI (H2-H3) reductions are shown, while we show only the 2015 IRDIS SDI (K1-K2) reduction. We also show a weighted combination of the IFS SDI images for each epoch. Each IFS wavelength channel was weighted by the average flux predicted by the best fitting \cite{2012ApJ...756..172M} and \cite{2015ApJ...804L..17T} models. The strong detection of the companion in H2 and the lack of a H3 counterpart indicates the presence of strong methane absorption. All images have the same scale and orientation, and were generated by removing 20 PCA modes.}
  \label{fig:imaging_figure}
\end{figure*}

The PCA-reduced IRDIS and IFS images are shown in Fig. \ref{fig:imaging_figure}, for reductions removing 20 PCA modes. We clearly detect a companion, that we refer to as HD\,4113C, and use its observed position to optimize and inform the remaining data analysis steps.

An additional SDI reduction was performed for both the IFS and IRDIS datasets. For IRDIS, the images in the H3 filter were rescaled by the ratio of the wavelengths using an FFT-based approach, and subtracted from the H2 images taken simultaneously. Due to instrumental throughput and stellar flux differences between the two channels, the values in the H3 images were scaled to have the same total flux in an annular region of the speckle halo centred on the edge of the AO-correction radius. The same PCA algorithm was then performed on the resulting images.

For each IFS image, the corresponding images in other wavelength channels were rescaled by the ratio of the wavelengths and flux-scaled using the same approach used for IRDIS. A PSF reference was constructed from the median of these rescaled images, and subtracted from the original image. Using the known position of the brown dwarf from the IRDIS reduction, only wavelength channels where the companion position was shifted by 1.5\,FWHM or more were considered to minimize self-subtraction. This was repeated for each image at each wavelength, before applying PCA.

\begin{figure}
  \centering
  \includegraphics[width=\columnwidth]{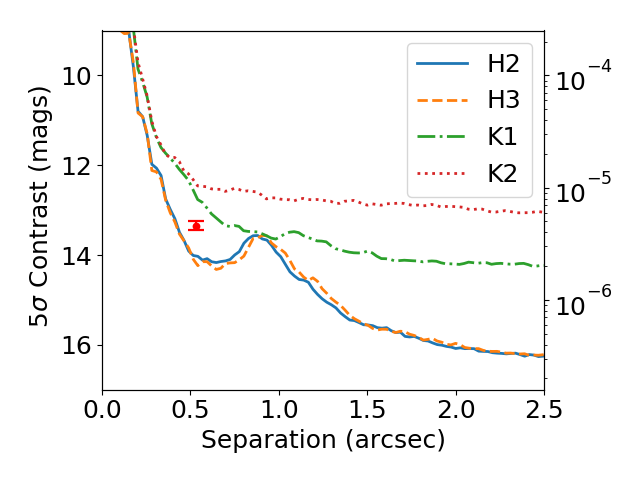}
  \caption{The $5\sigma$ contrast obtained for the four IRDIS datasets (K1 and K2 from the 2015 data, and H2 and H3 from the 2016 data). The measured separation and contrast of the companion detected in the H2 band is marked in red.}
  \label{fig:imaging_contrast}
\end{figure}

An estimation of the noise as a function of radius from the star was obtained from the standard deviation of the pixel values measured in an annulus of width $\lambda/D$. The location of HD\,4113C was masked out for this calculation to avoid biasing the measurements. This was then converted to a contrast curve accounting for the throughput of the PCA reduction and small sample statistics \citep[following the approach of][]{2014ApJ...792...97M}. The 5$\sigma$ contrast curves obtained through this method are shown in Fig \ref{fig:imaging_contrast}.

The position and relative flux of HD\,4113C were measured from the IRDIS data by subtracting a shifted and flux-scaled copy of the observed PSF of HD\,4113A from the final combined image. The PSF was obtained using the mean of the unobscured images of HD\,4113A taken with the ND filter. The separation, position angle and flux of the subtracted PSF were varied to minimize the residuals in a small ($3\lambda/D$) area. This was achieved using a nonlinear least-squares optimizer. A true north offset of $-1.75\pm0.08$\,deg and a plate scale of $12.255\pm0.009$\,mas/pix were used to calculate the separation and position angle of the companion, using the measurements presented in \cite{2016arXiv160906681M}. We also adopt a systematic uncertainty in the star position of $\pm 3$\,mas.

To estimate the uncertainties on the fitted position and flux, we applied a bootstrapping procedure. The PCA-subtracted and derotated frames were saved before stacking to generate a final image. We resampled the data (with replacement) to generate 1000 sets of frames, with each set having the same number of frames as the original data. These were collapsed to give 1000 final images. The fitting process described above was then performed on each, giving a distribution of values for the separation, position angle and contrast. The mean and standard deviation of these distributions were used as an estimate of the parameter values and their uncertainties.

As an additional measure, this procedure was then repeated for a range of PCA reductions by varying the number of modes, protection angle and annulus width. The scatter between the mean values for each reduction were included in our listed uncertainties.

For the IFS data, a different approach was taken to measure the companion flux, due to the low signal-to-noise ratio or non-detection of HD\,4113C in many spectral channels. The position of the companion was fixed to that measured from the stacked image, and a fit was performed to the flux only. The uncertainty was estimated by performing the same procedure at positions both in a ring with a radius of 2 FWHM around HD\,4113C and at a range of position angles at the same separation. The scatter of these values was taken as the uncertainty.


\begin{figure*}
  \centering
    \begin{subfigure}{0.9\textwidth}
        \includegraphics[width=0.9\textwidth]{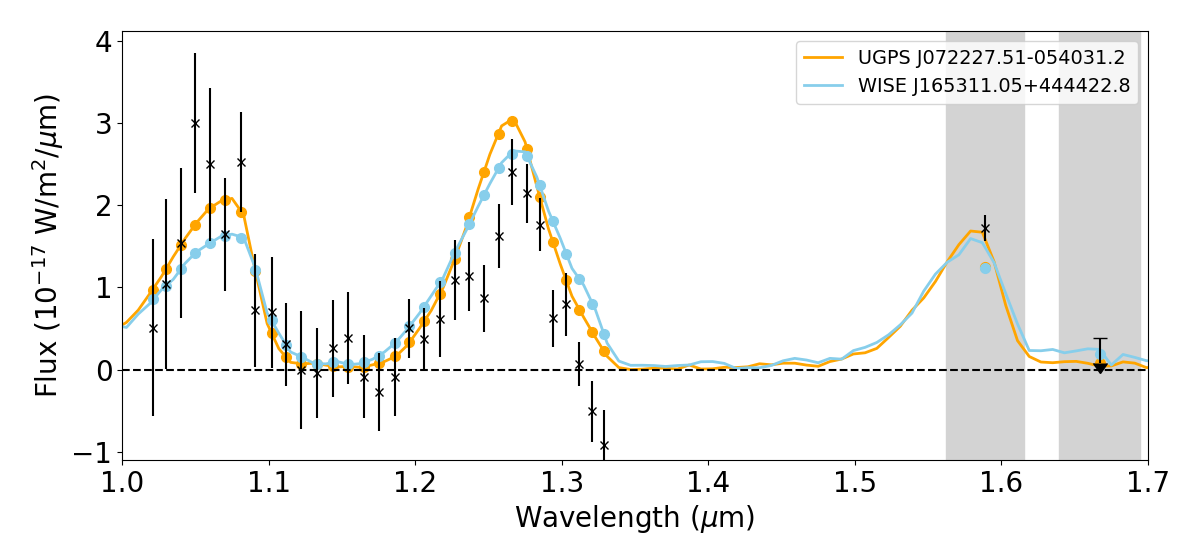}
        \caption{The measured spectrum of HD\,4113C extracted from the IFS and IRDIS images. Two peaks are seen in the IFS spectrum at 1.07\,$\mu$m and 1.27\,$\mu$m, which correspond well with the spectral features seen in late T dwarfs. The two best-fitting spectra from the SpeX Prism library are overplotted, with flux scaled to match HD\,4113C. The predicted flux values for each datapoint are shown with filled circles while the datapoints are marked with crosses. The FWHM of the IRDIS filters are shown with grey bars. Fitting was performed to the IFS and IRDIS data simultaneously.}
        \label{fig:imaging_spectrum_spex}
    \end{subfigure}
    
    \begin{subfigure}{0.9\textwidth}
        \includegraphics[width=0.9\textwidth]{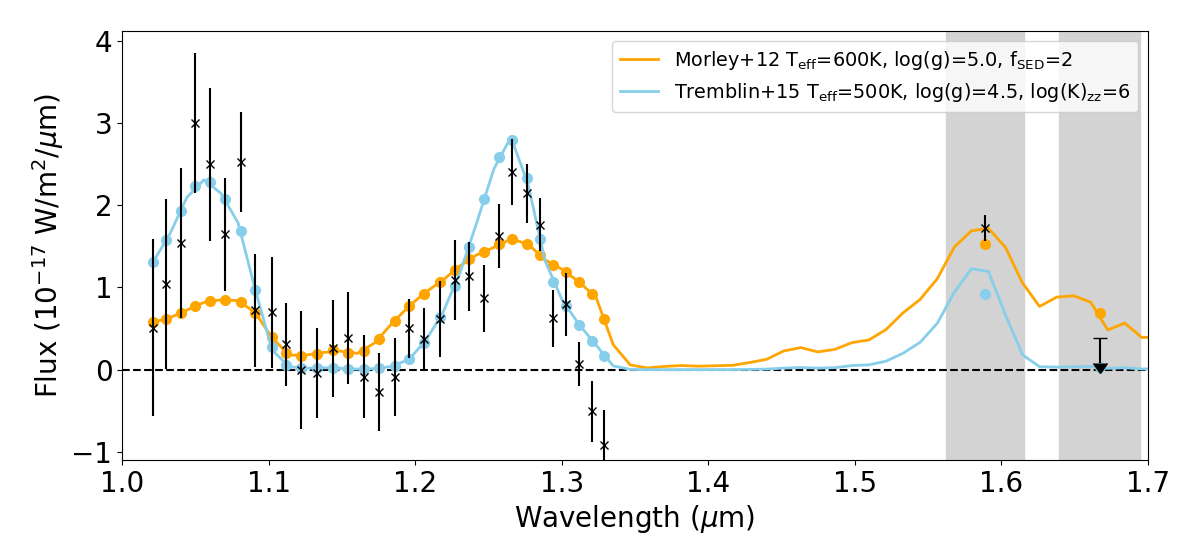}
        \caption{Same as Fig. \ref{fig:imaging_spectrum_spex}, this time showing the comparison between the observed spectrum and the best-fit spectra from the \cite{2012ApJ...756..172M} and \cite{2015ApJ...804L..17T} models, using radii of 1.4 and 1.5 \rjup\, respectively}
        \label{fig:imaging_spectrum_model}
    \end{subfigure}

  \caption{Comparison of the observed spectrum of HD\,4113C with similar ultracool dwarfs and brown dwarf atmospheric models.}
  \label{fig:imaging_spectrum}
\end{figure*}

\subsection{Archival SPHERE data}
We also re-reduced the SPHERE observations of HD\,4113 published in \cite{2017arXiv170105664M} (Program 096.C-0249, PI: Moutou), taken on 2015-10-08. These observations were taken in IRDIFS-EXT mode following a similar observing strategy to our observations described above, and are summarized in Table \ref{tab:observing_log}. The IRDIS data were taken in the K1-K2 mode ($\lambda_{K1} = 2.1025$\,$\mu$m, $\lambda_{K2} = 2.2255$\,$\mu$m), while the IFS data cover a range of wavelengths from 0.97-1.66\,$\mu$m with an average spectral resolution of 34.5. These data include a maximum field rotation of 51 deg between coronagraphic frames.

Our data reduction procedure was identical to that of the IRDIFS data described in the previous section.

\section{Results\label{sec:results}}
\subsection{SPHERE 2016 high-contrast imaging data}

HD\,4113C was detected at high significance in both IRDIS H2 and IRDIS SDI reductions, but was not recovered in the H3 filter, suggesting the presence of strong methane absorption. The measured relative photometry and astrometry are given in Table \ref{tab:imaging_results}, while the IRDIS PSF-subtracted images are shown in Fig. \ref{fig:imaging_figure}. We adopt a lower limit on the contrast of 14.8\,mag for the H3 filter, using the 3$\sigma$ contrast curve at the location of the H2 detection.

In the IFS data utilizing PCA only, no significant detection was found in any individual wavelength channel or in the mean taken over wavelength.

After applying SDI to the IFS data, the companion is clearly detected, with 14 of the wavelength channels showing an excess of flux above $1.5\sigma$ significance at the location of the IRDIS detection. A weighted mean image is shown in Fig. \ref{fig:imaging_figure}. The data reduction for wavelengths below 1.02$\mu$m suffer from considerably more residual noise in the SDI+PCA reduction, and are not included in the subsequent analysis.

The measured contrast ratios from IFS and IRDIS were converted to apparent flux measurements through multiplication by a BT-NextGen spectral model \citep{2012RSPTA.370.2765A} with T$_{\text{eff}}$=5700\,K, [Fe/H]=0.3\,dex and $\log g$=4.5, close to the values determined in Section \ref{stellarParams}. The model spectrum was flux-scaled using the distance and radius of HD\,4113A. The extracted spectrum is shown in Fig. \ref{fig:imaging_spectrum}.

\begin{table*}
\centering
\caption{The measured astrometry and photometry of HD\,4113C from the SPHERE observations.}
\begin{tabular}{ccccccc}
Instrument & Filter & Date & $\rho$ (arcsec) & $\theta$ (deg) & Contrast (mag) & SNR (peak) \\
\hline
IRDIS & H2 & 2016-07-20 & 0.535 $\pm$ 0.003 & 41.3 $\pm$ 0.4 & 13.35 $\pm$ 0.10 & 12\\
IRDIS & H3 & 2016-07-20 &   & & >14.8 & \\
IRDIS & H2-H3 & 2016-07-20 & 0.533 $\pm$ 0.003 & 42.0 $\pm$ 0.4 & 13.5 $\pm$ 0.2 &  16\\
IFS & YJ & 2016-07-20 & 0.528 $\pm$ 0.005 & 41.1 $\pm$ 0.4 & & 7.6 \\
IFS & YH & 2015-10-08 & 0.513 $\pm$ 0.006 & 41.2 $\pm$ 0.6 & & 6.2
\end{tabular}
\label{tab:imaging_results}
\end{table*}

\begin{figure}
  \centering
  \includegraphics[width=\columnwidth]{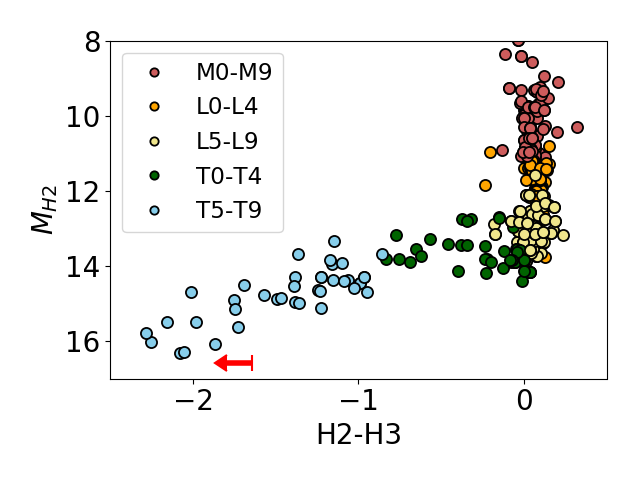}
  \caption{A color-magnitude diagram showing the predicted H2-H3 colors and H2 absolute magnitudes for objects in the SpeX Prism Library. The flux of HD\,4113C and the $3\sigma$ upper limit on its H2-H3 color are shown with a red arrow. Its position is compatible with objects at the end of the T sequence, showing deep methane absorption indicative of T and Y dwarfs.}
  \label{fig:imaging_color_mag}
\end{figure}

\subsection{SPHERE archival high-contrast imaging data}

After processing the 2015 data, we were unable to unambiguously recover the companion HD\,4113C in the IRDIS images. Close to the location of the 2016 detection, we find an excess of flux with 2-3$\sigma$ significance in the SDI reduction. However, due to the lower SNR, our reduction of the IRDIS data does not provide substantial constraints on the position or flux of the companion and we have excluded it from the following analysis.

In the IFS SDI reduction, we also find an excess of flux in wavelength channels corresponding to the J and H band peaks found in the 2016 data, again with lower significance. We find 6 channels at which the companion is visible at SNR$>$4, and by combining the final images with knowledge of the measured spectrum from the 2016 data, we obtain a detection with SNR$>$5. From this image we were able to calculate the position of HD\,4113C listed in Table \ref{tab:imaging_results}. Due to the lower SNR, we do not include the flux measurements in our analysis of the companion.

The difference in the observed position between the two epochs demonstrates orbital motion at the 3$\sigma$ level. The observed motion also differs strongly from that expected by a background object due to the 125\,mas\,yr$^{-1}$ proper motion of HD\,4113, demonstrating that it is indeed co-moving.

The detection limits obtained from the IRDIS data are shown in Fig.~\ref{fig:imaging_contrast}. Our IFS reduction produces contrast limits similar to those reported in \cite{2017arXiv170105664M}.

\subsection{Companion Properties} \label{sec:companion_properties}
To allow a consistent comparison with other objects and to estimate the mass of HD\,4113C, we converted the H2 flux to an absolute magnitude using the distance of HD\,4113 from Table \ref{tab-mcmc-StellarParams}.

In addition to the H2 measurement, we used the IFS spectrum to calculate a synthetic J band magnitude for HD\,4113C using a top-hat function between $1.125-1.365$\,$\mu$m. For both the J and H2 filters, we used the calibrated spectrum of Vega from \cite{2007ASPC..364..315B} to calculate the flux corresponding to zero magnitude. The resulting absolute photometric magnitudes are listed in Table \ref{tab:photometry_mass}.

The measured absolute photometry can then be compared with substellar evolutionary models. We used the COND \citep{2003A&A...402..701B} substellar isochrones to predict the companion photometry as a function of age, temperature and mass in each band. We then interpolated this grid to estimate the parameters of HD\,4113C.

The measured H2 magnitude $M_{H2}=16.6\pm0.1$ corresponds to a mass of $36\pm5$\mjup. The isochrones predict a temperature of $690\pm20$\,K and an H3 contrast of 14.9\,mag, which is close to our lower limit of 14.8\,mag. The J band magnitude $M_{J}=18.5\pm0.2$ leads to estimates of $29\pm5$\mjup\, and $580\pm20$\,K. However, few benchmark T dwarfs with measured mass, temperature and luminosity are available with age estimates similar to HD 4113 and so we caution that these values are in a relatively unconstrained region of the COND grid and may be unreliable.

\begin{table}
\centering
\caption{The measured absolute photometry of HD 4113\,C.}
\begin{tabular}{cccccc}
Band & App. Mag & Abs. Mag & COND Mass & Temp. (K) \\
\hline
H2 & $19.7\pm0.1$ & $16.6\pm0.1$ & $36\pm5$ & $690\pm20$\\
J & $21.6\pm0.2$ & $18.5\pm0.2$ & $29\pm5$ & $580\pm20$\\
\end{tabular}
\label{tab:photometry_mass}
\end{table}

In order to estimate the spectral type of HD\,4113C, we used the SpeX Prism library of near-IR spectra of brown dwarfs \citep{2014ASInC..11....7B} using the {\it splat} python package \citep{2016AAS...22743408B}. Each spectrum was flux calibrated to the distance of HD\,4113. To compare the goodness of fit of each spectrum ($F_{k}$) to the observed combined IFS and IRDIS spectrum ($f$), we use the $G_k$ statistic used by \cite{2008ApJ...678.1372C} and others, where each measurement is weighted by its spectral FWHM:
\begin{equation}
    G_k = \sum_{i=1}^{n} w_i \left( \frac{f_i - C_k F_{k,i}}{\sigma_i} \right)
\end{equation}
with the weights given by $w_i = \Delta \lambda_i / \sum_{j=1}^{n} \Delta \lambda_j$. $C_k$ is a constant used to scale the flux of each spectrum, chosen to minimize $G_k$. We investigated the best fits with and without the inclusion of the $C_k$ factor, to isolate the effects of the spectral shape and the absolute flux.

Each SpeX Prism spectrum was converted to the appropriate spectral resolution of each IFS measurement by convolution with a Gaussian. The FWHM used for the convolution was assumed to be 1.5$\times$ the separation between each wavelength channel. To predict the IRDIS fluxes, we used the transmission curves for the H2 and H3 filters.

Comparison with all objects with known distances in the SpeX Prism library shows that HD\,4113C is most compatible with the coldest objects in the library, with spectral types of T8-T9 providing the best fit. Without including the $C_k$ factor to scale the flux of the SpeX Prism spectra, the individual flux values are marginally lower than any object in the library, leading to a stronger preference for late T spectral types. This same trend is seen when considering the IFS and IRDIS results separately. In all cases, the only T9 object in the library gave the best fit to the spectrum, suggesting a most likely spectral type of T9.

The best fits to the observed spectrum are given by the objects UGPS J072227.51-054031.2 \citep[$G_k = 2.86$ ]{2010MNRAS.408L..56L} and WISE J165311.05+444422.8 \citep[$G_k = 3.22$]{2011ApJS..197...19K}, field dwarfs of type T9 and T8 respectively. A comparison of the measured spectrum with those of these two objects is shown in Fig. \ref{fig:imaging_spectrum_spex}.

Using the predicted H2 and H3 fluxes from the SpeX Prism Library, we produced a color-magnitude diagram in Fig. \ref{fig:imaging_color_mag}. The observed H2 flux is similar to the faintest objects in the Library, adding additional weight to the idea that the companion is close to the T/Y transition. The $3\sigma$ upper limit on the H2-H3 color is consistent with spectral types T7 and later.

To further constrain the physical properties of HD\,4113C, we compared the observed SED to synthetic spectra for cool brown dwarfs from \cite{2012ApJ...756..172M}, \cite{2014ApJ...787...78M} and \cite{2015ApJ...804L..17T} using the same minimization technique as was used for the SpeX Prism Library. Each set of models gives fluxes as measured at the surface of the object, and can be flux-scaled to the distance of HD 4113 after assuming a radius. We considered radii between 0.5-1.5\rjup\, when comparing the models.

The \cite{2012ApJ...756..172M} models include the effects of sulfide clouds at low temperatures and cover a range of effective temperatures $T_\text{eff}$ = 400-1300\,K and surface gravities  $\log g$ = 4-5.5, with condensate sedimentation efficiency coefficients f$_{\mathrm{SED}}$ = 2-5.
The \cite{2014ApJ...787...78M} models extend the previous models to lower temperatures through the inclusion of water ice clouds and partly cloudy atmospheres. They cover a range of effective temperatures $T_\text{eff}$ = 200-450\,K, surface gravities $\log g$ = 3-5 and condensate sedimentation efficiency coefficients f$_{\mathrm{SED}}$ = 3-7, and assume 50\% of the surface is covered by clouds.
The \cite{2015ApJ...804L..17T} models include the effects of convection and non-equilibrium chemistry and cover a range of effective temperatures $T_\text{eff}$ = 200-1000\,K and surface gravities $\log g$ = 4-5, with vertical mixing parameters $\log K_{zz}$ = 6-8. The parameters of the best fitting models are shown in Table \ref{tab:spectral_fitting_results}.

For the \cite{2012ApJ...756..172M} models, lower temperature spectra fit the observed values more closely, with the best fitting model being $T_\text{eff}$=600\,K, $\log g=5.0$, f$_{\mathrm{SED}}$=2, $R=1.4$\,\rjup. The model spectra are not strongly affected by changes in f$_{\mathrm{SED}}$ in this regime of temperature and $\log g$, and for all f$_{\mathrm{SED}}$ values in the grid, the same effective temperature and surface gravity provide the best fit. These models provide a reasonable fit to the data, but predict a higher H3 flux than our upper limits, and appear to underestimate the Y band flux.

The \cite{2014ApJ...787...78M} models did not produce a good match to the observed data. The best fitting model corresponds to $T_\text{eff}$=450\,K, $\log g=5$, f$_{\mathrm{SED}}$=5 and $R=1.5$\,\rjup. These parameters are at the edge of the model grid, and the $G_k$ value was significantly higher than the other model sets. They failed to simultaneously fit the fluxes of the J and H band peaks with a radius $R<1.5$\,\rjup.

When comparing the data with the \cite{2015ApJ...804L..17T} model spectra, we find best fit parameters of $T_\text{eff}$ = 500\,K, $\log g$ = 4.5, $\log K_{zz}$ = 6 and $R=1.5$\,\rjup. Similar to the previous model set, these models have difficulty reproducing the fluxes of the J and H peaks while constrained to $R<1.5$\,\rjup.

The radius and temperature are highly correlated for each set of models (since $T_\text{eff} \propto R^{-0.5}$ for a black-body). Lower temperature models provide a better fit to the shape of the spectrum, but fail to reproduce the flux.
When fixing the radius to 1\,\rjup, the best fit temperatures are 700\,K for the \cite{2012ApJ...756..172M} grid and 600\,K for the \cite{2015ApJ...804L..17T} grid, with $G_k$ values of 3.2 and 5.1 respectively. 

If the restriction on the radius is removed, the fits converge to temperatures of 300\,K and 325\,K for the \cite{2015ApJ...804L..17T} and \cite{2014ApJ...787...78M} models with $G_k$ values of 0.94 and 2.7 respectively, while the \cite{2012ApJ...756..172M} models prefer the lower edge of the model grid at 400\,K with a $G_k$ value of 1.6. These fits match the shape of the observed spectrum, but do not correspond to physically reasonable radii.

The best-fitting spectrum from the \cite{2012ApJ...756..172M} and \cite{2015ApJ...804L..17T} model grids are shown in Fig. \ref{fig:imaging_spectrum_model}.

Due to the relatively low SNR of the spectrum, the lack of K and L photometry, and the use of only a single epoch of data, we caution that more work is needed before accurate limits can be placed on the effective temperature, radius and surface gravity measurements.

\begin{table*}
\centering
\caption{Comparison of best fitting model parameters extracted from the spectral fitting.}
\begin{tabular}{ccccccc}
Model & $T_\text{eff}$(K) & $\log g$(cgs) & f$_{\mathrm{SED}}$ & $\log K_{zz}$ & $R$(\rjup) & $G_k$ \\ \hline
\cite{2012ApJ...756..172M} & 600 & 5.0 & 2 & & 1.4 & 3.5 \\
\cite{2014ApJ...787...78M} & 450 & 5.0 & 5 & & 1.5 & 8.3 \\
\cite{2015ApJ...804L..17T} & 500 & 4.5 & & 6 & 1.5 &  3.5\\
\end{tabular}
\label{tab:spectral_fitting_results}
\end{table*}

\subsection{Orbit determination and dynamical masses}
\label{sec:orbit_fit}

\begin{figure}
  \centering
    \includegraphics[width=0.95\columnwidth]{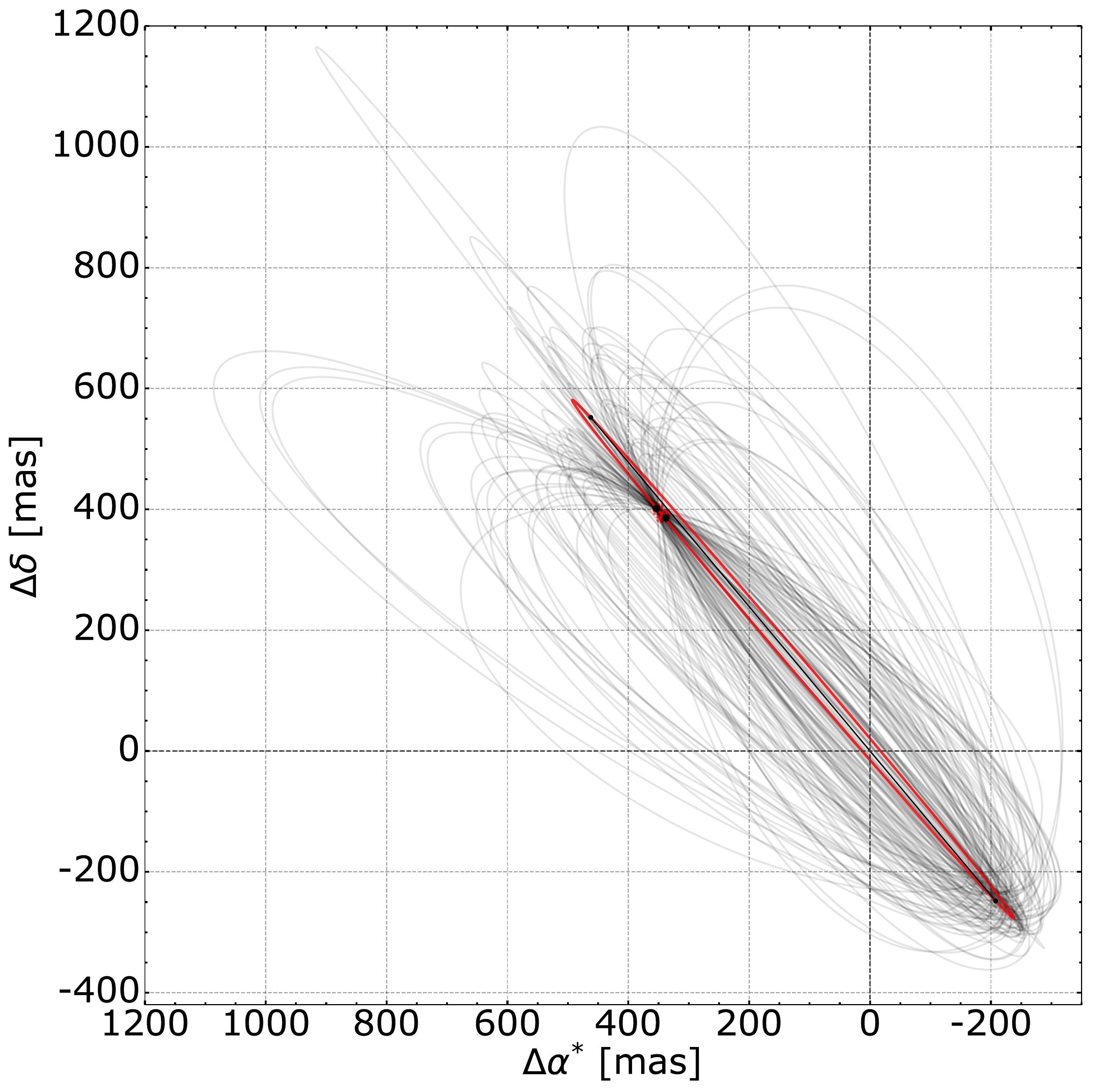}
    \includegraphics[width=0.95\columnwidth]{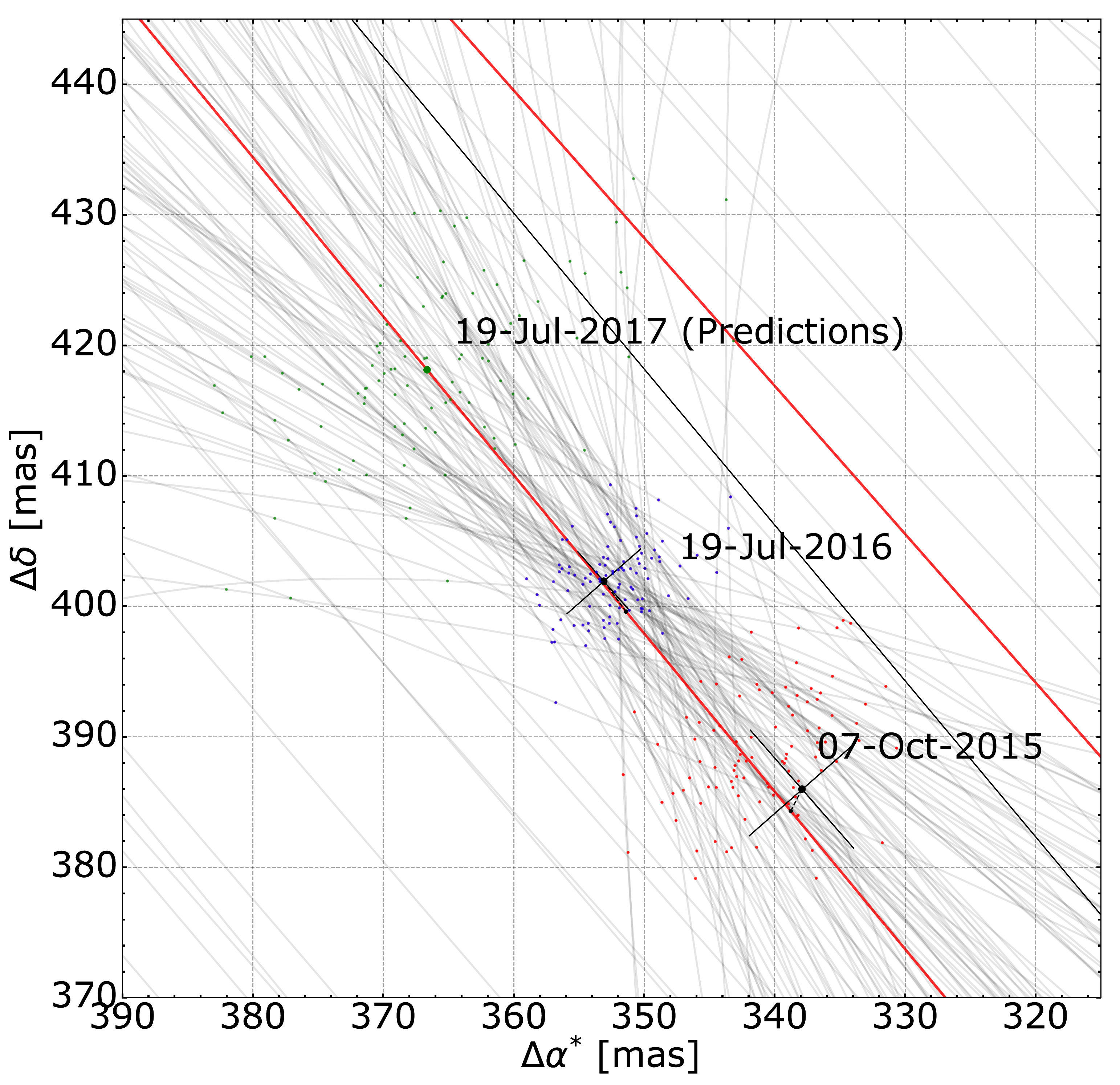}
 \caption{Observed and predicted orbital motion of HD\,4113 AC, shown as the relative projected position of HD\,4113C. A series of orbits are plotted, drawn from the posterior distribution of the combined RV+imaging analysis. The bold red curve corresponds to the Maximum Likelihood solution. The bottom figure is a zoom of the orbit, centered on the date of the observations.}
  \label{fig:BD}
\end{figure}

To constrain the orbital parameters of the planet and brown dwarf, we performed a combined fit to the radial velocity and direct imaging data. For this purpose, we use the 2016 IRDIS H2 astrometry and the astrometry extracted from the 2015 IFS YH combined image, both reported in Table \ref{tab:imaging_results}.

The initial RV data analysis presented in this paper was accomplished using a set of online tools hosted by the Data \& Analysis Center for Exoplanets (DACE)\footnote{The DACE platform is available at http://dace.unige.ch while the online tools to analyse radial velocity data can be found in the section Observations$=>$Radial Velocities} which performs a multiple Keplerian adjustment to the data as described in \citet{2016A&A...590A.134D}. It allows us to derive initial conditions for a combined analysis of the direct imaging and RV data, performed within a Bayesian formalism. 

To model the observed data, a Keplerian is assigned to the inner planet ($P=526.57$\,d, $m\sin{i}=1.66$\,M$_{\rm Jup}$) which is only seen in the RV measurements. A second Keplerian is added to reproduce both the observed RV drift and the positions of the imaged brown dwarf companion.

The observed signal is modeled with two Keplerians and five RV offsets (one offset for each RV instrument). We model the noise using a nuisance parameter for each RV instrument \citep{Gregory-2005}. These nuisance parameters are quadratically added to the individual measurement errors. We combine the signal and the noise model into a likelihood function that is probed using  {\it emcee} - a python implementation of the Affine Invariant Markov chain Monte Carlo (MCMC) Ensemble sampler  \citep{2013PASP..125..306F,Goodman-2010}. The full technical description of the combined data analysis is given in Appendix \ref{appendixB} and we focus here on the data analysis results.

Despite possessing only two direct imaging measurements and the partial RV coverage of the orbit, we are able to bring significant constraints on the geometry of the orbit and mass of HD\,4113\,C. Indeed, the observed RV drift displays a large peak-to-peak value of $\sim$365 \ms\, over the 18~yrs of CORALIE observations, setting a clear minimum mass boundary for the brown dwarf. 

Within this framework and based on the MCMC posterior distribution, we are able to set confidence intervals for the orbital elements and physical parameters of HD\,4113\,C. At the 1$\sigma$ level, the period ranges between 87\,yr and 134\,yr, its semi-major axis, between 20.3~AU and 27.1~AU while its mass is in the range 61-71~\mjup. The eccentricity is well constrained with a value ranging between $e=0.31$ and $e=0.46$ at 1$\sigma$.

\begin{figure}
  \centering
  \includegraphics[width=\columnwidth]{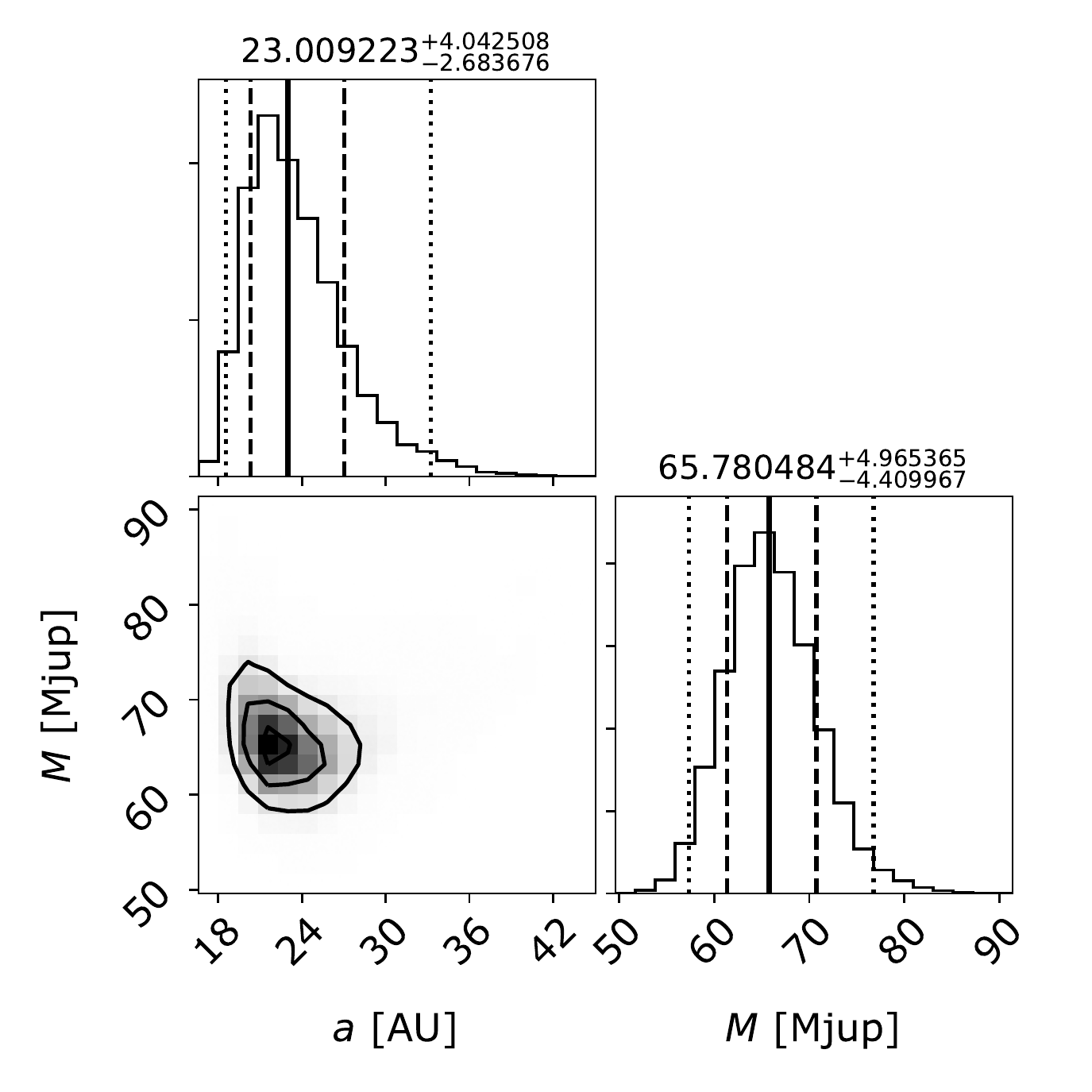}
  \caption{Marginalized 1-D and 2-D posterior distributions of the Mass vs. Semi-Major axis corresponding to the global fit of the RV and direct imaging models. Confidence intervals at $2.275\%, 15.85\%, 50.0\%, 84.15\%, 97.725\%$ are over-plotted on the 1-D posterior distributions while the median  $\pm$ 1 $\sigma$ value are given on top of each 1-D distribution. 1, 2 and 3 $\sigma$ contour levels are over-plotted on the 2-D posterior distribution. }
  \label{fig:mass-separation}
\end{figure}

The results of the MCMC analysis are summarized in Table~\ref{tab-mcmc-PhysicalParams} and the corresponding orbits are shown in Fig. \ref{fig:BD}. A mass-separation diagram based on the marginalized 2D posterior distribution is represented in Fig~\ref{fig:mass-separation}.

\begin{table*}
\caption{Orbital elements and companion properties derived from the posterior distribution of the two-Keplerian Model. The mean longitude, $\lambda_{0}=M_{0}+\omega+\Omega$, is given at the reference epoch of BJD=55500.0\,d.}     
\label{tab-mcmc-PhysicalParams}

\begin{tabular}{lc|cc|cccc}     
&&{ \bf HD4113Ab} & &\multicolumn{4}{c}{ \bf HD4113C}   \\
Param&Units&Value(mode)&(Err-Err+)& Value(median)&(Err-Err+)&CI(2.275)&CI(97.725)  \\
\hline  
$P$            &               &   526.586\,[d] &(    -0.016,     0.013)&   104.6\,[yr]&(   -17.9,    29.0)&    75.4&   180.6\\ 
$K$            &[\ms]          &    89.16&(    -0.58,     1.18)&   387.9&(   -33.7,    42.7)&   325.6&   484.2\\ 
$e$            &               &     0.8999&(    -0.0016,     0.0020)&     0.377&(    -0.062,     0.081)&0.260&     0.543\\
$\omega$       &[deg]          &   -48.45&(    -0.77,     0.87)&   200.3&(   -12.2,    10.3)&   176.4&   219.5\\ 
$T_{0}$        &[d]            & 55642.738&(    -0.083,     0.090)& 48127&(  -800.,   690.)& 46518& 49384\\ 
$\Omega$       &[deg]          &-&-&    41.8&(    -7.2,     7.2)&    27.3&    56.5\\ 
$\lambda_{0}$  &[deg]          &    49.11&(    -0.71,     0.86)&   -46.9&(   -17.2,    14.4)&   -80.2&   -20.0\\ 
$i$            &[deg]          &-&-&    89.1&(   -12.0,    12.3)&    65.2&   112.9\\ 
$m.\sin{(i)}$  &[$M_{\rm Jup}$]&     1.602&(    -0.075,     0.076)&    64.4&(    -4.5,     5.0)&    55.7&    74.8\\ 
$m$            &[$M_{\rm Jup}$]&-&-&    65.8&(    -4.4,     5.0)&    57.4&    76.8\\ 
$a_{r}$        &[AU]           &     1.298&(    -0.030,     0.030)&    23.0&(    -2.7,     4.0)&    18.5&    33.2\\ 
\end{tabular}
\end{table*}

\subsection{Dynamics and formation}
\label{sec:dynamics}

As seen in Table \ref{tab-mcmc-PhysicalParams}, the eccentricity of \object{HD~4113}Ab is particularly high ($e_\mathrm{Ab} \approx 0.9$). The Lidov-Kozai mechanism \citep[see][]{lidov_evolution_1962,kozai_secular_1962} has often been proposed to explain highly eccentric planetary orbits.
In the case of HD\,4113A\,b, this idea was raised at the time of its discovery by \citet{2008A&A...480L..33T}. The observed radial velocity drift was evidence of a high mass, long-period companion that would naturally interact gravitationally with the planet.

Using the orbital parameters derived from the new radial velocity and direct imaging measurements, we are able to test the Lidov-Kozai scenario in more detail. We adopt the best-fitting (maximum likelihood) parameters for the planet and brown dwarf (see Table~\ref{tab-mcmc-Allparams}) in order to have a coherent set of orbital elements.
The only parameters that are unconstrained by the observations are the inclination ($i_\mathrm{Ab}$)
and the longitude of the ascending node ($\Omega_\mathrm{Ab}$) of the planet.
We thus let these parameters vary.
We also let the mass of the planet ($M_\mathrm{Ab}$) vary accordingly with its inclination,
since the radial velocity data constrain only its minimum mass ($M_\mathrm{Ab}\sin i_\mathrm{Ab}$).
We run a set of 1681 ($41\times 41$) simulations of the system using
the GENGA integrator \citep[see][]{grimm_genga_2014},
with $i_\mathrm{Ab}$ varying in the range $[5^\circ,175^\circ]$ ($M_\mathrm{Ab} \in [1.67-19.21]$),
and $\Delta\Omega = \Omega_\mathrm{Ab}-\Omega_\mathrm{C}$ varying in the range $[-180^\circ,180^\circ]$.
The time step is set to 0.001 yr and the orbits are integrated for 1 Myr,
including the effects of General Relativity.

\begin{figure}
  \centering
  \includegraphics[width=\linewidth]{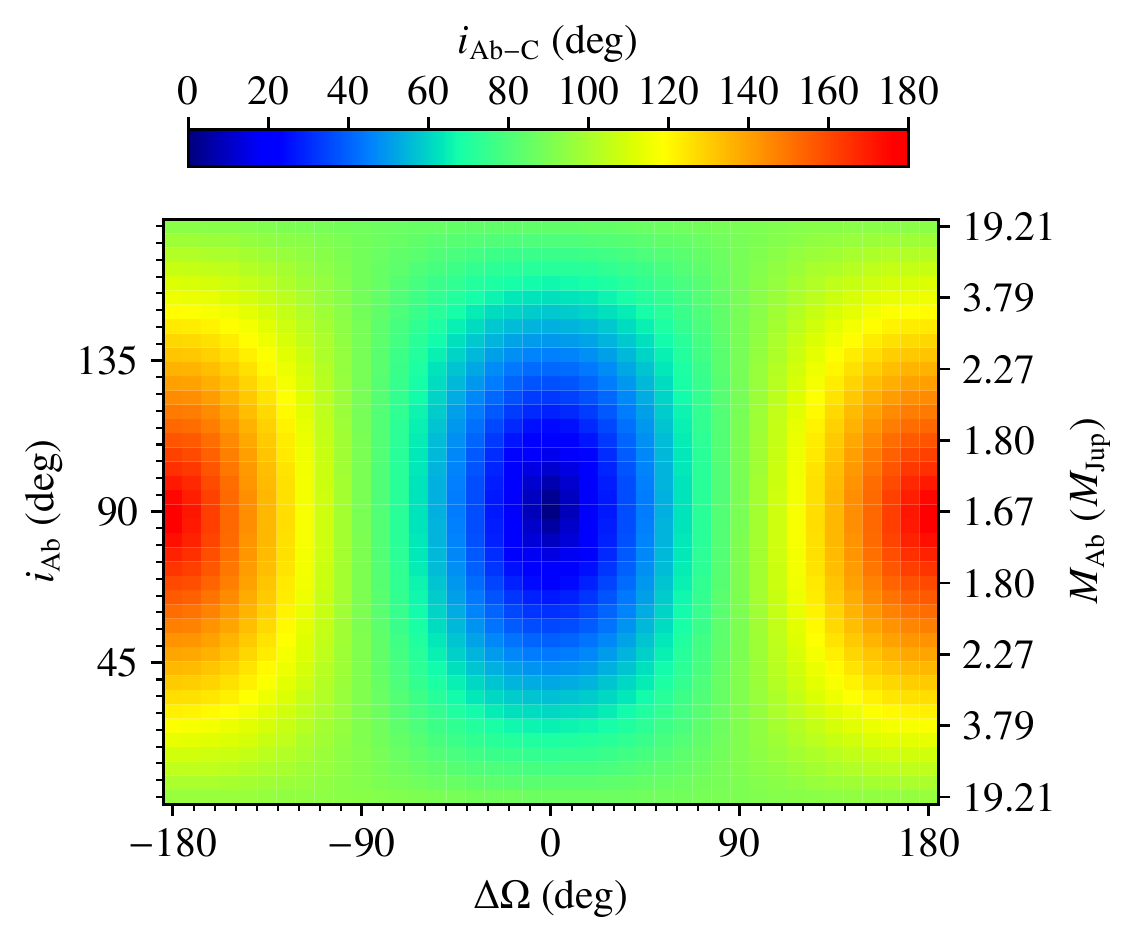}
  \caption{Initial mutual inclination between the planet
  and brown dwarf orbits ($i_\mathrm{Ab-C}$),
  as a function of the planet's inclination ($i_\mathrm{Ab}$),
  and longitude of the ascending node
  ($\Delta\Omega = \Omega_\mathrm{Ab}-\Omega_\mathrm{C}$).}
  \label{fig:DynI}
\end{figure}

\begin{figure}
  \centering
  \includegraphics[width=\linewidth]{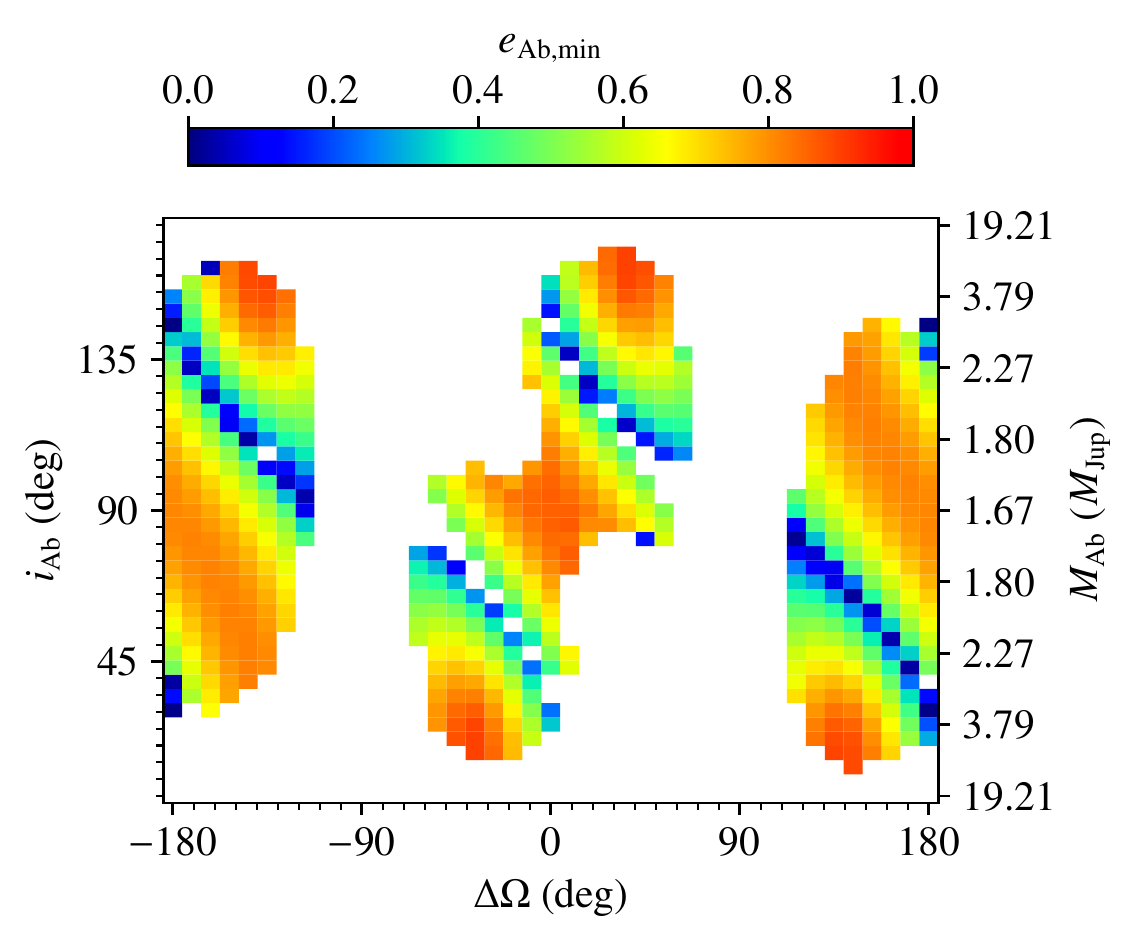}
  \caption{Minimum value reached by the planet's eccentricity
  during the 1 Myr of integration ($e_\mathrm{Ab,min}$),
  as a function of the planet's initial inclination ($i_\mathrm{Ab}$),
  and longitude of the ascending node
  ($\Delta\Omega = \Omega_\mathrm{Ab}-\Omega_\mathrm{C}$).
  White points correspond to unstable simulations,
  that stopped before 1 Myr.}
  \label{fig:DynII}
\end{figure}

\begin{figure}
  \centering
  \includegraphics[width=\linewidth]{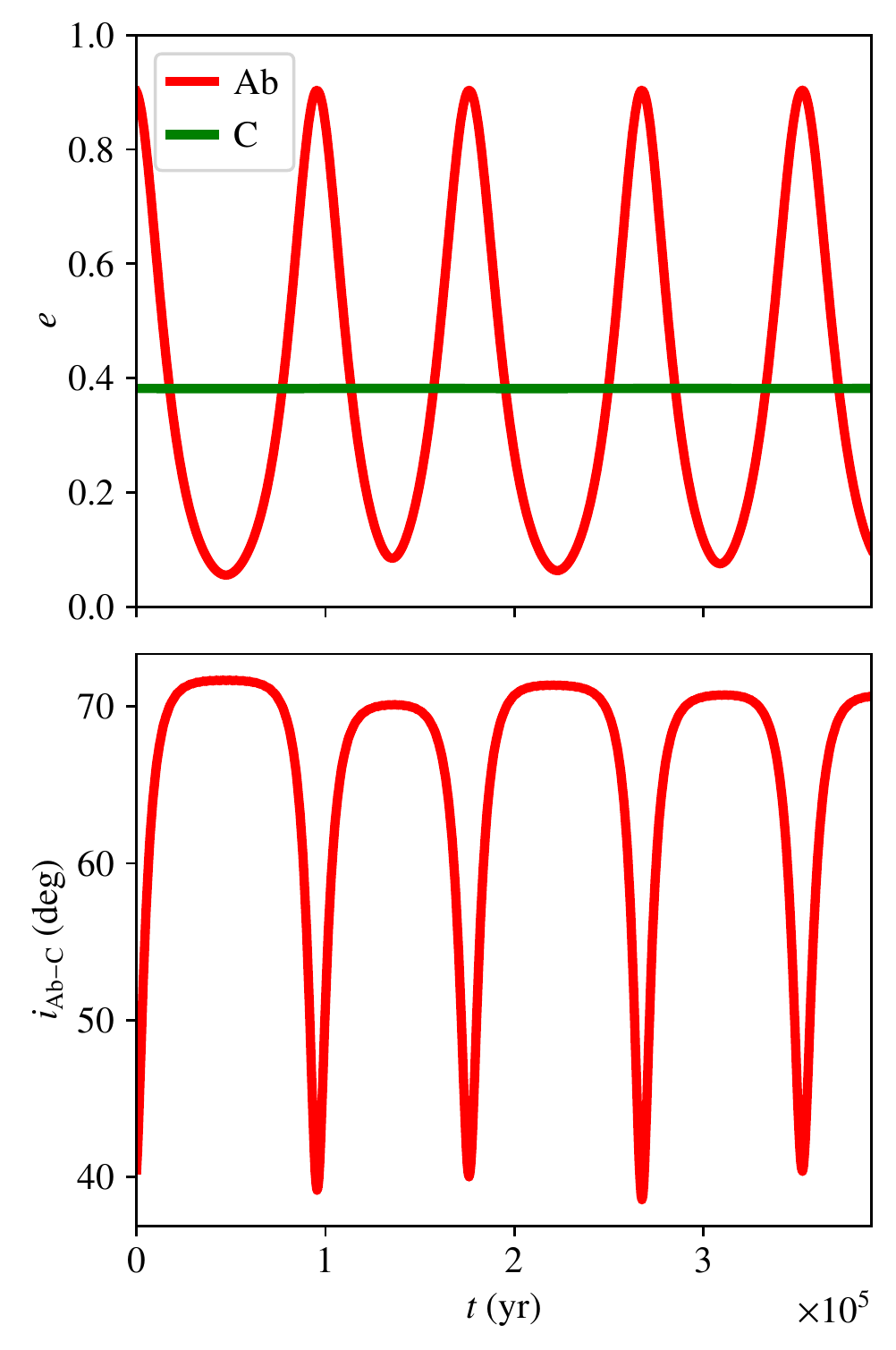}
  \caption{Example of temporal evolution of the eccentricities of the planet
  and the brown dwarf (\textit{top})
  as well as their mutual inclination (\textit{bottom}).
  The initial conditions are set using the best-fitting parameters (see Table~\ref{tab-mcmc-Allparams}),
  and $i_\mathrm{Ab} = 128.25^\circ$ ($M_\mathrm{Ab}\approx2.13$\,\mjup), and $\Delta\Omega=18^\circ$.
  This corresponds to an initial mutual inclination of about $40.47^\circ$.
  The minimum value of the planet's eccentricity is about 0.06,
  and is reached when the mutual inclination reaches about $72^\circ$.
  }
  \label{fig:DynIII}
\end{figure}

Figure~\ref{fig:DynI} shows the initial mutual inclination between the orbits of the planet and brown dwarf ($i_\mathrm{Ab-C}$), plotted as a function of the two unknown parameters. 
Figure~\ref{fig:DynII} shows the minimum value reached by the planet's eccentricity
during the whole simulation.
Unstable integrations are stopped and drawn in white in Fig.~\ref{fig:DynII}.
In particular, the simulations in which the planet approaches the star by less than 0.005 AU
(close to one stellar radius) are stopped.
Depending on the initial parameters ($i_\mathrm{Ab}$, $\Omega_\mathrm{Ab}$), strong Lidov-Kozai cycles
are observed and the eccentricity of the planet can reach very low values
(blue points in Fig.~\ref{fig:DynII}).
Figure~\ref{fig:DynIII} shows an example of a simulation where strong Lidov-Kozai cycles are observed.
Thus, it is possible that the planet formed on a quasi-circular orbit and that the observed
eccentricity (0.9) is the result of Lidov-Kozai oscillations with the observed brown dwarf.

From Fig.~\ref{fig:DynII} we could derive constraints
on the missing orbital elements of the planet
($i_\mathrm{Ab}$, $\Omega_\mathrm{Ab}$)
for the system to be stable (i.e. not located in a white region),
and for the planet to have formed with a low eccentricity (blue parts).
We note that several effects could modify the results shown in Fig.~\ref{fig:DynII}.
Considerable uncertainties remain on several of the orbital parameters for the brown dwarf, and changes in these orbital parameters would reflect on Fig.~\ref{fig:DynII}.
Moreover, in some of the simulations shown here,
the eccentricity of the planet reaches very high values.
In this case, the time step we use (0.001 yr)
is not sufficiently small to correctly model the periastron passage,
and some orbits that are detected as unstable
(white parts in Fig.~\ref{fig:DynII}) might be stable (and vice versa).
In addition, in the very high eccentricity cases,
tides may play an important role.
Indeed the combination of the Lidov-Kozai mechanism and tides
can induce an inward migration of the planet
\citep[see][]{wu_planet_2003,fabrycky_shrinking_2007,correia_tidal_2011}.
In this scenario, the planet initially forms
at a larger distance to the star and with a quasi-circular orbit.
It then undergoes strong Lidov-Kozai oscillations, which brings the planet very close to the star at periastron.
In addition to shrinking the planet orbit (decreasing semi-major axis),
tides (and general relativity) tend to damp the amplitude of the Kozai oscillations.
The maximum eccentricity reached during a Lidov-Kozai cycle does not change much,
but the minimum eccentricity increases.
When the Lidov-Kozai amplitude is vanishing,
the eccentricity is simply damped from a very high value, down to zero.
If HD\,4113A\,b is following this path,
the amplitude of the Lidov-Kozai cycles could have already been significantly damped.
Therefore, the minimum eccentricity reached by the planet during a Lidov-Kozai cycle
could be high,
even if the planet initially formed on a quasi-circular orbit.


\section{Discussion and Conclusion\label{sec:conclusion}}
The HD\,4113 system is an important laboratory for studying a range of issues in stellar and substellar physics. The primary star, HD\,4113A, hosts a planetary companion on a highly eccentric orbit. By fitting to 28 years of radial velocity data, we update the orbital parameters from those presented at the time of its discovery by \cite{2008A&A...480L..33T}. In addition, we present high-contrast imaging data that reveals the nature of a long period RV signal as originating from a cool brown dwarf companion.

The imaging data shows that the brown dwarf has a contrast in the SPHERE H2 filter of 13.35\,mag, and a deep methane absorption feature that causes it to have a H2-H3 color of $>1.45$\,mag. Using the COND isochrones, the H2 and synthetic J band absolute fluxes are consistent with masses of $36\pm5$\,\mjup\, and $29\pm5$\,\mjup\, respectively.

Fitting to the observed spectrum from IFS and IRDIS allows us to estimate the effective temperature, spectral type, radius and surface gravity of HD\,4113C. Through comparison with the SpeX Prism Library of brown dwarf spectra, we find that it is most consistent with the coldest objects in the library, with the best match coming from a T9 field dwarf. Additional support for a late-T spectral type comes from the position of HD\,4113C on the H2-H3 color-magnitude diagram (Fig. \ref{fig:imaging_color_mag}).

Using the atmospheric models of \cite{2012ApJ...756..172M} and \cite{2015ApJ...804L..17T}, we find best fit values of $T_{\text{eff}}$ = 500-600\,K, $\log g=$4.5-5 and $R$=1.4-1.5\,\rjup.

However, due to the reliance on a single epoch, moderate SNR spectrum, additional high SNR data covering a wider range of wavelengths are needed to further characterize HD\,4113C, and to confirm the spectral type, effective temperature, radius and surface gravity estimates. Indeed the radius estimate is much higher than predicted for old, high-mass brown dwarfs, suggesting that more work may be needed to understand the system parameters. In particular, additional spectra to constrain the shape of the J and H band peaks, and photometric measurements at longer wavelengths (K and L band) would help in this task.

We place constraints on the orbital and physical parameters of HD\,4113C through orbital fitting to the radial velocity and imaging data. This suggests a mass of $66^{+5}_{-4}$\,\mjup, and a moderate eccentricity of $0.38^{+0.08}_{-0.06}$. With longer time baselines for both the imaging and radial velocity data, orbital fitting will provide tighter constraints on the mass, and this system could act as a useful mass-luminosity benchmark for late T dwarfs.

When comparing the derived properties of HD\,4113C to those of other directly imaged companions to main sequence stars, we find a strong correspondence with GJ\,504b \citep{2013ApJ...774...11K} in many aspects. The late-T spectral type and J band absolute magnitude ($\sim$18.7\,mag) in particular suggest that GJ\,504b is a useful analog. Both objects also orbit metal enriched, solar-like stars, with recent data suggesting that GJ\,504 may also have a solar-like age of $\sim4$\,Gyr \citep{2015ApJ...806..163F}.

In addition to GJ\,504b, HD\,4113C joins a growing list of cool, T dwarf companions imaged around main sequence stars, including the T4.5-T6 planet 51 Eri b \citep{2015Sci...350...64M} and the T8 brown dwarf GJ 758 B \citep{2009ApJ...707L.123T}.

HD\,4113C also joins HD 19467B \citep{2014ApJ...781...29C} as the second imaged T dwarf with a measured radial velocity acceleration that allows it to be used as a benchmark object. While HD 19467B appears to have a higher effective temperature of $\sim1000$\,K, it's dynamical mass and age are similar to that of HD\,4113C.

Through dynamical simulations we show that the observed high eccentricity (e$\sim$0.9) of the giant planet HD\,4113A\,b could be the result of strong Lidov-Kozai cycles caused by interactions with the orbit of the imaged brown dwarf. For certain ranges of the unknown orbital parameters $i_{Ab}$ and $\Delta\Omega$, eccentricity oscillations of >0.8 are observed, typically with an oscillation period of $10^{4}-10^{5}$\,yr. This raises the possibility that HD\,4113A\,b formed on an initially quasi-circular orbit and subsequently underwent eccentricity excitation through interaction with HD\,4113C.

However, further questions about the formation of HD\,4113A\,b remain. The brown dwarf companion would be expected to truncate the circumstellar disk of HD\,4113A at 1/2 - 1/3 of its semi-major axis \citep{1994ApJ...421..651A}, which corresponds to 7-11\,AU on its present orbit. Forming a planetary companion in such an active dynamical system within a truncated disk may prove challenging.

An exciting prospect for further characterization of this system are the future Gaia data releases. The astrometric signal of the giant planet HD\,4113A\,b should be detectable with Gaia, and will give an accurate inclination measurement. The planet will introduce astrometric motion with a peak-to-peak amplitude of at least 0.08\,mas. In addition, a long term drift corresponding to the HD\,4113A-C orbit should also be visible. With a 10\,yr dataset, this would introduce a 13\,mas drift in the position of HD 4113A, with a maximum deviation of 0.8\,mas from linearity. Even with the current 3\,yr time baseline, the curvature induced by HD\,4113C should have an amplitude of $\sim0.1$\,mas.

The combination of Gaia astrometry with future RV and imaging datasets will allow for much tighter constraints on the orbits and physical parameters of each component. Most critically, Gaia astrometry will give a first measurement of the unknown orbital parameters of the planet. This will be crucial for studying the Lidov-Kozai interactions and stability of the system, by significantly reducing the parameter range to explore and by resolving the degeneracy with the unknown planet mass. Moreover, the combination of more accurate mass measurements of the brown dwarf with improved spectroscopy and photometry will be useful for comparison with atmospheric and evolutionary models.

While the imaging data does confirm the existence of a wide brown dwarf-mass companion seen in the RVs, the observed MCMC posterior for the mass of HD\,4113C shows an apparent underlying inconsistency between the input data. As seen in Fig. \ref{fig:mass-separation}, the predicted mass is close to the stellar/substellar boundary, while the spectrum obtained through direct imaging shows an extremely cool temperature that would suggest a much lower mass.

This combination of the dynamical mass, age and low temperature of the companion presents a challenge to brown dwarf cooling models. The COND models predict that a 66\,\mjup\, object would have cooled to $1200\pm170$K at $5.0^{+1.3}_{-1.7}$\,Gyr, significantly higher than the $500-600$\,K estimates from the best-fit spectral models. 

However, there are several scenarios that may explain this apparent disagreement. Most notably, the newly detected companion may be an unresolved brown dwarf binary system, leading to an increased dynamical mass and radius estimate without greatly influencing the shape of the observed spectrum. In the case of an equal mass binary, this would relax the tension between the temperature and radius in the spectral model fits, allowing for a pair of 500-600K objects with radii of 1\rjup\, to provide a good match to the observed data. This is close to the prediction of $640\pm80$\,K from the COND models, based on a 33\,\mjup\, object at $5.0^{+1.3}_{-1.7}$\,Gyr. In this case, given the lack of extended structure visible in the companion PSF, we can only place an upper limit of 2\,AU on the current projected separation of such a system. More stringent constraints may be given by considering the stability of the system as a whole.

An alternative possibility is that the presence of an additional companion to HD\,4113A could have biased the RV data and caused an overestimate of the dynamical mass.
For example, the long-term curvature in the RVs may be caused by an unseen planetary or brown dwarf companion at a smaller separation, while the imaged companion may be a longer period intermediate mass brown dwarf that contributes only a linear drift to the observed radial velocities.

We investigated this possibility numerically in our orbit fitting through the addition of a third body (``HD\,4113D'') at intermediate periods ($25$~[yr]$<P<$100~[yr]) which is not detected by SPHERE and which absorbs most of the RV drift.

Repeating the MCMC analysis results in a wide range of solutions due to the modest constraints of the observations with respect to the complexity of the model. However, we are able to extract from the posterior distribution a sub-sample of good solutions with a significantly lower mass for the brown dwarf imaged by SPHERE. In Table~ \ref{tab-mcmc-PhysicalParams-3keplerian} we show one example with a dynamical mass of $40$\,\mjup\, for the imaged companion, similar to the isochronal mass estimates from the SPHERE data.
 
While the present data do not provide any way to distinguish between this scenario and the 2 Keplerian model, additional imaging, radial velocity and astrometric data will allow this idea to be investigated further. With several imaging datasets taken over a much longer time period, the orbit calculated from the imaging data alone could be compared to that expected from the RV data. Any inconsistencies in the estimated parameters would then indicate that an additional body may be present in the system.

However, the 3 Keplerian scenario fails to explain the higher than expected flux (or equivalently, radius) of the imaged companion given its estimated effective temperature. Of these hypotheses, the binary brown dwarf scenario presents a more compelling case due to its capacity to explain all of the properties that have been explored in this paper, and the clearer dynamical stability of the system as a whole. Despite this, given the current data and the orbital coverage more work is needed to firmly establish the underlying cause of the mass discrepancy.

\begin{table}
\centering
\caption{An example of consistent orbital elements and companion properties for a three-Keplerian model compatible with our observations. The listed solution corresponds to one realization of a similar MCMC procedure to that used to produce Table \ref{tab-mcmc-PhysicalParams} and was chosen according to its probability (among the 1\% best solutions) and the low dynamical mass of the imaged brown dwarf.}
\label{tab-mcmc-PhysicalParams-3keplerian} 
\begin{tabular}{lc|cc|cccc}     
     &     & & { \bf potential} & \\
     &     &{ \bf HD\,4113Ab} & { \bf HD\,4113D} &{ \bf HD\,4113C}   \\
Param&Units&                &                &  \\
\hline  
$P$            &               &   526.590 [d]& 26.9 [yr]&    209[yr] \\ 
$K$            &[\ms]          &   90.02 &143 & 185\\ 
$e$            &               &    0.90& 0.43& 0.07\\ 
$\omega$       &[deg]          &    -47.57 &133.7 & 65.9 \\ 
$T_{RV_{MAX}}$  &[d]            &   -  & 58082  &69043 \\ 
$\Omega$       &[deg]          &      -&- &42.13 \\ 
$i$            &[deg]          &  -    &- &89.20 \\ 
$m.\sin{(i)}$  &[\mjup]&  1.59   &14 &40.31 \\ 
$m$            &[\mjup]&   -   &- &40.32 \\ 
$a_{r}$        &[AU]           &  1.29  &9.1 &36.0 \\ 
\vspace{1mm} 
\end{tabular}
\end{table}

\begin{acknowledgements}
This work has been carried out within the frame of the National Centre for Competence in Research ``PlanetS'' supported by the Swiss National Science Foundation (SNSF).\\
This publication makes use of The Data \& Analysis Center for Exoplanets (DACE), which is a facility based at the University of Geneva (CH) dedicated to extrasolar planets data visualisation, exchange and analysis. DACE is a platform of the Swiss National Centre of Competence in Research (NCCR) PlanetS, federating the Swiss expertise in Exoplanet research. The DACE platform is available at https://dace.unige.ch.\\
SPHERE is an instrument designed and built by a consortium consisting of IPAG (Grenoble, France), MPIA (Heidelberg, Germany), LAM (Marseille, France), LESIA (Paris, France), Laboratoire Lagrange (Nice, France), INAF - Osservatorio di Padova (Italy), Observatoire Astronomique de l'Université de Genève (Switzerland), ETH Zurich (Switzerland), NOVA (Netherlands), ONERA (France) and ASTRON (Netherlands) in collaboration with ESO. SPHERE was funded by ESO, with additional contributions from CNRS (France), MPIA (Germany), INAF (Italy), FINES (Switzerland) and NOVA (Netherlands). SPHERE also received funding from the European Commission Sixth and Seventh Framework Programmes as part of the Optical Infrared Coordination Network for Astronomy (OPTICON) under grant number RII3-Ct-2004-001566 for FP6 (2004-2008), grant number 226604 for FP7 (2009-2012) and grant number 312430 for FP7 (2013-2016). \\
We acknowledge financial support from the Programme National de Planétologie (PNP) and the Programme National de Physique Stellaire (PNPS) of CNRS-INSU. This work has also been supported by a grant from the French Labex OSUG@2020 (Investissements d’avenir – ANR10 LABX56). \\
This work has made use of data from the European Space Agency (ESA)
mission {\it Gaia} (\url{http://www.cosmos.esa.int/gaia}), processed by
the {\it Gaia} Data Processing and Analysis Consortium (DPAC,
\url{http://www.cosmos.esa.int/web/gaia/dpac/consortium}). Funding
for the DPAC has been provided by national institutions, in particular
the institutions participating in the {\it Gaia} Multilateral Agreement.\\
This publication makes use of data products from the Two Micron All Sky Survey, which is a joint project of the University of Massachusetts and the Infrared Processing and Analysis Center/California Institute of Technology, funded by the National Aeronautics and Space Administration and the National Science Foundation.\\
J.H. is supported by the Swiss National Science Foundation (SNSF), and the French National Research agency through the GIPSE grant ANR-14-CE33-0018. \\
This research has benefitted from the SpeX Prism Spectral Libraries, maintained by Adam Burgasser at http://pono.ucsd.edu/~adam/browndwarfs/spexprism

\end{acknowledgements}

\bibliographystyle{aa}

\bibliography{big_bibliography}

\clearpage

\begin{appendix}
\section{NACO L-band direct imaging data\label{appendixA}}
We present here the results of the 2013-08-21 data taken with NACO (Program 091.C-0721, PI S{\'e}gransan). The observing sequence consisted of 419 cubes of 128 frames, each with an exposure time of 0.2\,s. The maximum field rotation during the sequence was 112\,deg. This dataset was processed with the GRAPHIC pipeline \citep{2016MNRAS.455.2178H}, using a similar reduction sequence as the SPHERE data described in Section 3, without the application of SDI. Prior to applying PCA, the data were binned into 419 frames by taking the median over each cube.

\begin{figure}
  \centering
  \includegraphics[width=\columnwidth]{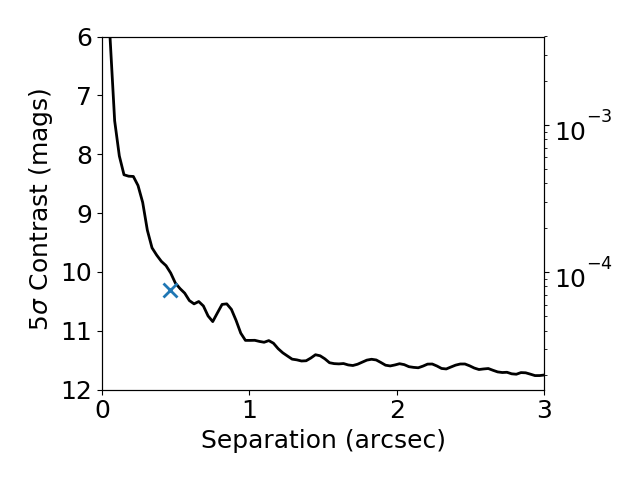}
  \includegraphics[width=0.8\columnwidth]{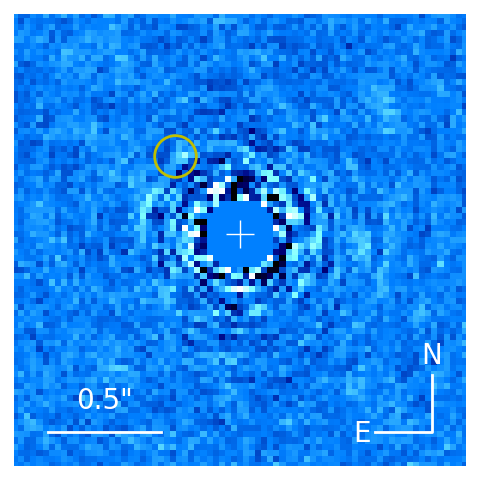}
  \caption{The $5\sigma$ contrast obtained from the NACO L' data taken in 2013. The separation and predicted contrast from the orbit fitting is marked with a blue cross. At this separation, the corresponding limit is 10.3\,mag. Despite the close correspondence between the predicted contrast and the $5\sigma$ limits, we do not unambiguously detect the companion. The final reduced image is also shown, after removing 150 PCA modes. The predicted position for the 2013 epoch is shown with a yellow circle. While the contrast curve has been corrected for companion self-subtraction introduced by the PCA procedure, the image has not. Through injection of fake companions, we estimate a 35\% flux loss for objects at the predicted separation of HD 4113\,C.}
  \label{fig:naco_saturated_detec_limits}
\end{figure}

The detection limits and final reduced image are presented in Fig. \ref{fig:naco_saturated_detec_limits}. The companion was not detected in this data, with the $5\sigma$ contrast limit at the location of the SPHERE detection being 10.3\,mag. At its predicted location based on the orbital fit in Sec. \ref{sec:orbit_fit}, the $5\sigma$ contrast limit is 10.0\,mag. Using the BHAC15 and COND isochrones, these correspond to mass limits of 73 \mjup.

Using the COND models, the expected L band contrast is 10.3\,mag using the dynamical mass of 66\,\mjup. The H2 and J band contrasts predict L band contrasts of 13.9 and 14.9\,mag respectively.

\section{MCMC Details\label{appendixB}}

\subsection{Likelihood function}

Throughout this section we assume that the observed signals are modeled with two independent Keplerians and five RV offsets corresponding to the number of RV instruments. The noise associated with each measurement is assumed to be independently drawn from a Normal distribution with zero mean $N(0,\sigma_{i})$. The observed signal $y_{i}$ is written in Eq.~\ref{eq1} as a sum between the model $f_{i}({\bf X})$ for parameters ${\bf X}$ and a realization of the noise $e_{i}$. $y_{i}$ represents both the measured radial velocities and the direct imaging data ( {\it i.e.} $\rho$, $PA$).
\begin{equation}
\label{eq1}
y_{i} = f_{i}({\bf X})+ e_{i}
\end{equation}

The likelihood function is given in Eq.~\ref{likelihood} where $\sigma_{i}$ is the estimated error on each data point $y_{i}$ and $\sigma_{0}$  is an additional white noise parameter \citep[see][]{Gregory-2005} added to each RV instrument. These nuisance parameters  account for any additional noise sources that were not included in the measurement's error bars. 
\begin{equation}
\label{likelihood}
\log{\mathcal{L}} = -\frac{1}{2} \sum_{i=1}^{N} \log{(\sigma_{0}^{2}  + \sigma_{i}^{2})} 
                    -\frac{1}{2} \sum_{i=1}^{N} \frac{ (y_{i}-f_{i}({\bf X}))^{2} } {\sigma_{0}^{2}  + \sigma_{i}^{2} }
\end{equation}

\subsection{Choice of parameters}
Our model for the RV measurements is composed of 5 linear terms: the systematic RV of HIRES ($\gamma_{HIRES}$), the offset between CORAVEL and HIRES ($\Delta V(CORA-HIRES)$), the offset between CORALIE-98 and HIRES ($\Delta V(C98-HIRES)$), the offset between CORALIE-07 and HIRES ($\Delta V(C98-HIRES)$) and the offset between CORALIE-14 and HIRES ($\Delta V(C14-HIRES)$). The CORALIE offsets originate from the different instrument upgrades performed over the 19 years of observations.

We model the Keplerian corresponding to HD\,4113A\,b with the parameters $\log{P}$, $\log{K}$, $e$, $\omega$ and $M_{0}$, using the period $P$, radial velocity semi-amplitude $K$, eccentricity $e$, argument of periastron $\omega$ and the mean anomaly $M_{0}$ at a given reference epoch ($t_{\rm ref}$=2 455 500.0 d).

We use a different set of variables to model HD\,4113C to reduce correlations between them. We use $\log{P}$,   $\sqrt{e}\,\cos{\omega}$,  $\sqrt{e}\,\sin{\omega}$, $m_{C}\sin{i}$,  $T_{MIN(rv)}$, $i$ and  $\Omega$. The last four variables correspond to the minimum mass of the brown dwarf, the date at which the minimum RV occurs, the orbital inclination and the longitude of the ascending node.

\subsection{Choice of priors}
All adjusted parameters have uniform priors with the exception of the RV offsets. We apply priors to the RV offsets between the CORALIE instruments, $\Delta V(C98-C07)=N(0,5)$ and
$\Delta V(C98-C07)=N(12,5)$ has been calibrated using RV standard stars that were observed with both C98 and C07, allowing us to adopt a Gaussian prior centered on zero with a standard deviation of 5 \ms. The prior on the offset between CORAVEL and CORALIE is a truncated normal distribution defined as $\Delta V(CORA-C07)=TN(-51,200,-200,200)$.

We also include a set of priors for the parallax and masses of the system HD\,4113AC based on our external knowledge. We adopted a Gaussian prior for the parallax based on the latest Gaia measurements and we choose a uniform prior for the mass of HD\,4113C with an upper  at half of the mass of the primary star.
For the mass of HD\,4113~A, we used
a split normal distribution with asymmetric standard deviation taken from Table~\ref{tab-mcmc-StellarParams}.

\subsection{MCMC runs}

We probed the model parameter space using  {\it emcee}, a  python implementation of the Affine Invariant Markov chain Monte Carlo (MCMC) Ensemble sampler  \citep{2013PASP..125..306F,Goodman-2010} and the likelihood function described in the above subsections.

We ran several MCMC simulations. For the first MCMC run, we performed 4\,500 iterations using 50 walkers with initial conditions drawn from the solution obtained using DACE. A first set of parameter posterior distributions were derived using a burn-in phase of 1\,500 iterations.
In order to obtain a statistically robust sample, we ran a second MCMC simulation with 100 walkers drawn from the posterior of our first simulation with a total of 800\,000 iterations which guarantees a faster convergence to equilibrium. 

We derive our final sample by :
\begin{itemize}
    \item discarding 5\% of the walkers based on the properties of each walker posterior distribution (standard deviation, maximum value, median value).
    \item computing the correlation time scale ($\tau$) of each walker and discarding the ones with $\tau>500$. 
    \item eliminating the initialization bias by discarding the first $20 \tau$ elements of each walker \citep[following][]{zbMATH01124118}.
    \item sampling each walker according to their coherence time to build the final statistical sample. 
\end{itemize}

By doing so, we are left with a sample of $61\,200$ independent data points which allow  to retrieve and characterize the underlying parameter distributions with a $1/\sqrt{61200}\approx0.4\%$ accuracy \citep{zbMATH01124118} which is close to the 3$\sigma$ confidence intervals.

\begin{figure*}
  \centering
  \includegraphics[width=0.6\textwidth]{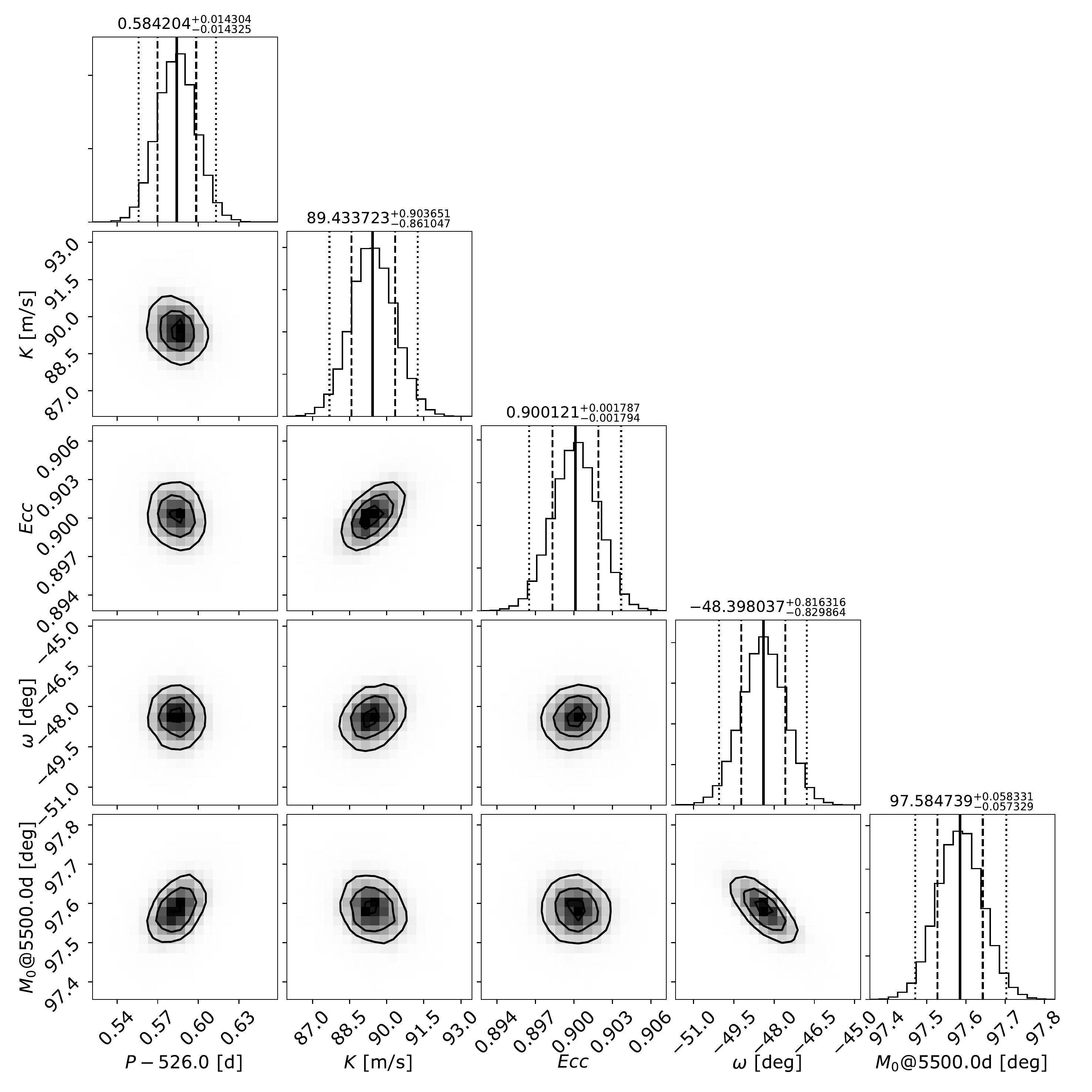}
  \caption{ Marginalized 1-D and 2-D posterior distributions of the model parameters corresponding to the global fit of the RV and direct imaging models for HD\,4113\,A\,b. Confidence intervals at $2.275\%, 15.85\%, 50.0\%, 84.15\%,9 7.725\%$ are over-plotted on the 1-D posterior distribution while the median  $\pm$ 1 $\sigma$ value are given on top of each 1-D posterior distributions. 1, 2 and 3 $\sigma$ contour levels are over-plotted on the 2-D posterior distributions. The model parameters adjusted during the MCMC run are shown ($P$, $K$, $e$, $\omega$, $M_{0}$)}
  \label{fig:Annex1}
\end{figure*}

\begin{figure*}
  \centering
  \includegraphics[width=\columnwidth]{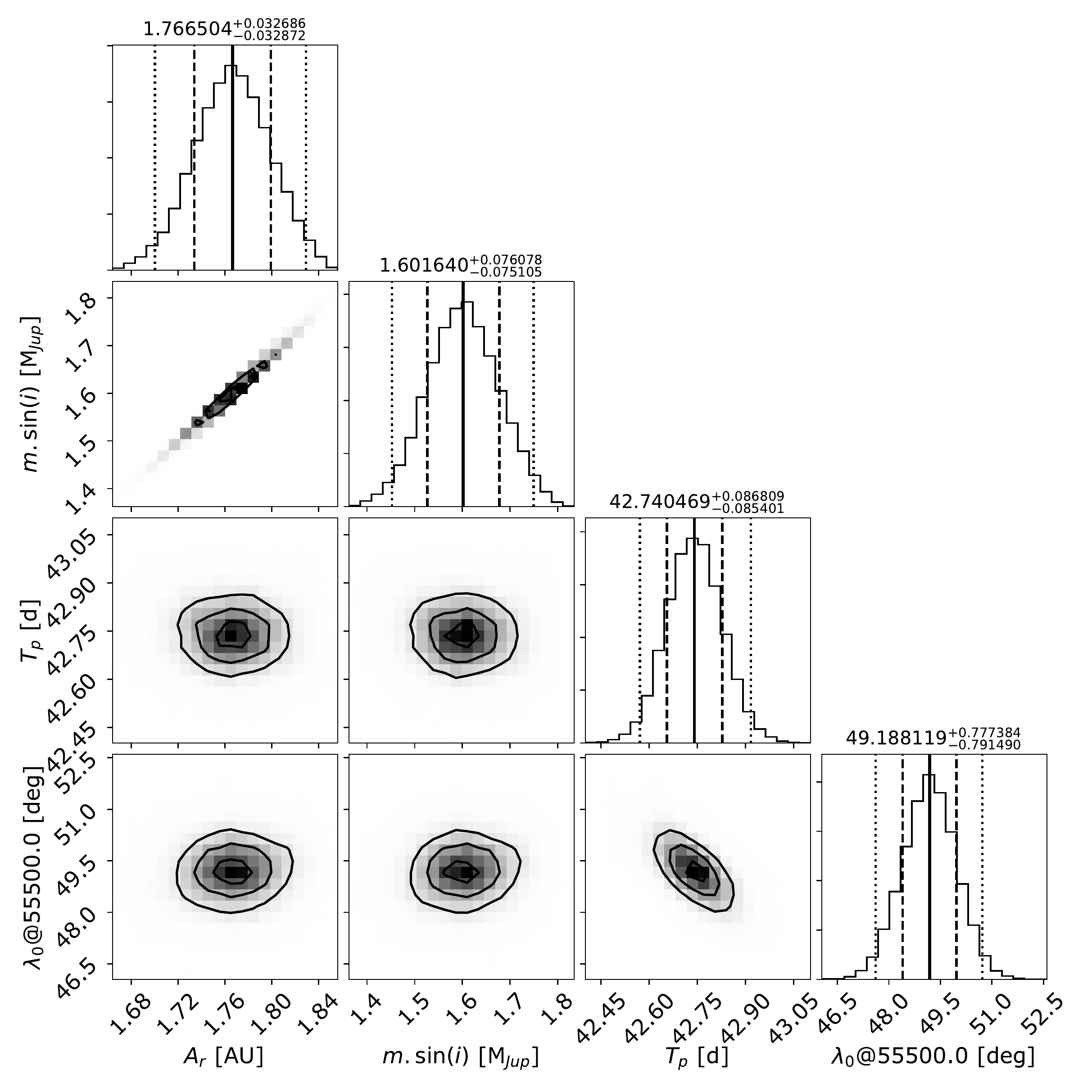}
  \caption{Same as Fig. \ref{fig:Annex1}, showing the additional parameters derived from the MCMC posterior samples ($a$, $m\sin{i}$, $T_{p}$, $\lambda_{0}$) for HD\,4113A\,b}
  \label{fig:Annex2}
\end{figure*}

Figures~\ref{fig:Annex1} and \ref{fig:Annex2} shows the marginalized 1-D and 2-D posterior distributions of the Keplerian model parameters of the giant planet HD\,4113\,A\,b. As illustrated, the distributions are close to Gaussian, with small linear correlations. The corresponding parameters and confidence intervals are listed in Table ~\ref{tab-mcmc-Allparams}.

\begin{figure*}
  \centering
  \includegraphics[width=0.97\textwidth]{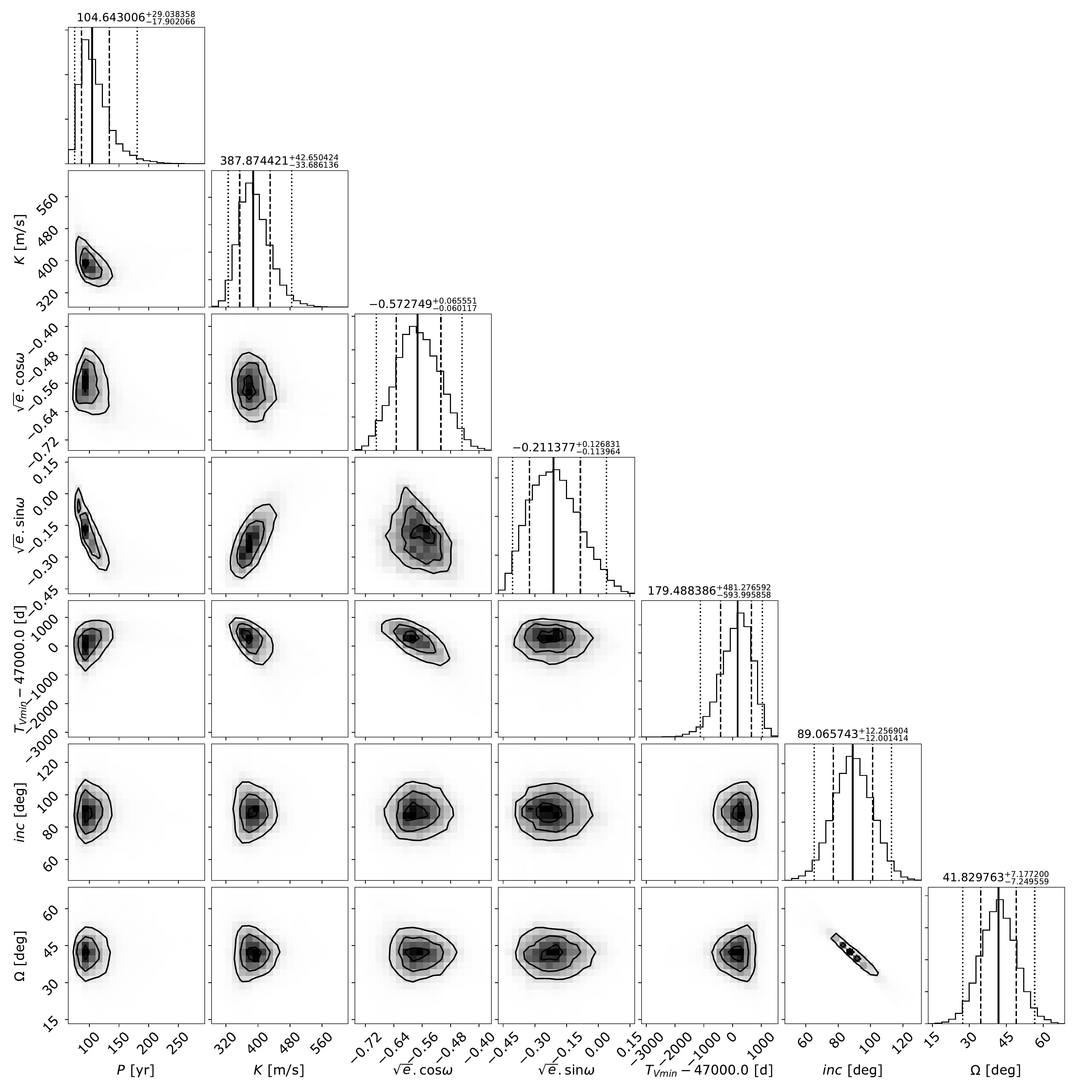}
  \caption{Same as Fig. \ref{fig:Annex1}, for HD\,4113C, showing the correlations and marginalized posterior distributions for the model parameters adjusted during the MCMC run ($P$, $K$, $\sqrt{e}\cos{\omega}$, $\sqrt{e}\sin{\omega}$, $T_{Vmin}$, $i$, $\Omega$) }
  \label{fig:Annex3}
\end{figure*}

\begin{figure*}
  \centering
  \includegraphics[width=0.97\textwidth]{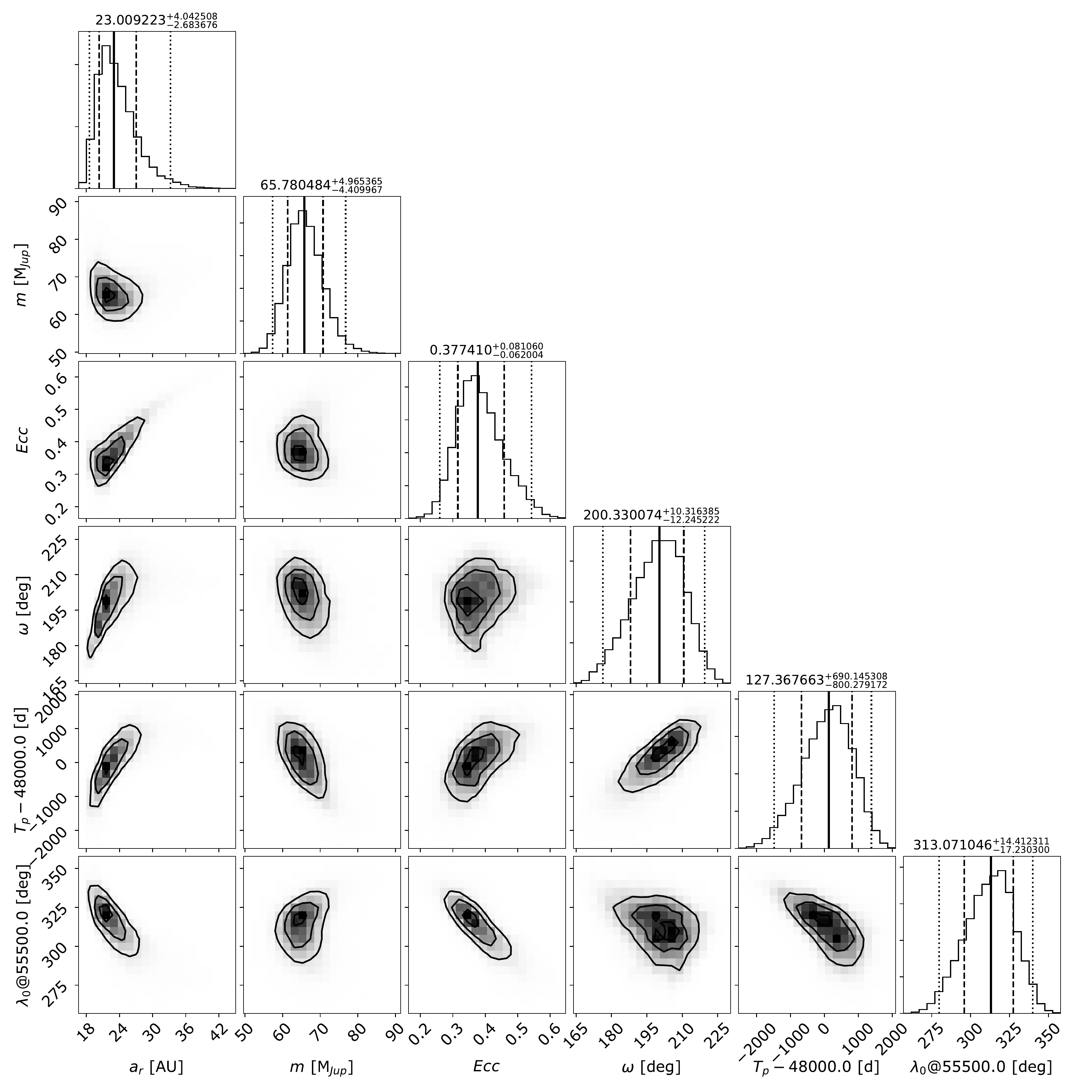}
  \caption{Same as Fig. \ref{fig:Annex2}\ref{fig:Annex2}, showing the additional parameters derived from the MCMC posterior samples for HD\,4113C ($a$, $m$, $e$, $\omega$, $T_{p}$, $\lambda_{0}$).}
  \label{fig:Annex4}
\end{figure*}

The case of HD\,4113\,C is less straightforward due to the presence of strong correlations in the 2-D posterior distributions (see Fig.~\ref{fig:Annex3} and ~\ref{fig:Annex4}). Furthermore, most of the 1-D distributions are non-Gaussian, which makes the use of mean value and error bars potentially misleading. Instead, we prefer to list the values corresponding to the maximum likelihood, the sample median  as well as the mode of the distribution in Table~\ref{tab-mcmc-Allparams}. Confidence intervals corresponding to $1\sigma$ and $2\sigma$ are also given for each parameter as well as the prior distribution used.

\begin{landscape}
\begin{table}
\tiny
\caption{Priors and parameters probed with the MCMC. Distributions:$U$: uniform, $N$: normal, $SN$:split normal, $TN$:truncated normal}  \label{tab-mcmc-Allparams}
\begin{tabular}{lcclcccccccc}
\hline
Param. & Units & Max(Like) & Med & Mod &Std & CI(15.85) & CI(84.15) &CI(2.275) & CI(97.725) & Prior\\
\hline
\multicolumn{11}{c}{ \bf Likelihood}\\
\hline
$\log{\rm Posterior}$&               &  -711.66&  -720.26&  -720.53&     2.91&  -723.87&  -717.32&  -728.22&  -714.93&               \\ 
$\log{\rm Likelihood}$&               &  -710.25&  -718.70&  -718.27&     2.75&  -722.13&  -715.95&  -726.28&  -713.73&               \\ 
$\log{\rm Prior}$&               &    -1.42&    -1.30&    -0.99&     0.92&    -2.57&    -0.57&    -4.33&    -0.21&               \\ 
\hline
\multicolumn{11}{c}{ \bf External priors}\\
\hline
$\pi$          &[mas]          &    24.094&    23.938&    23.807&     0.433&    23.454&    24.440&    22.976&    24.955&$N(23.986,\sqrt{0.237^2+0.43^2})$\\ 
$M_{A}$        &[M$_{\odot}$]  &     1.132&     1.050&     1.047&     0.063&     0.980&     1.125&     0.910&     1.196&$SN(1.05,-\sqrt{0.02^2+0.07^2},\sqrt{0.03^2+0.07^2})$\\ 
\hline
\multicolumn{11}{c}{ \bf RV offsets}  \\
\hline  
$\gamma(HIRES)$&[\ms]          &     -58.0&   -78.9&   -79.5&    46.7&  -130.6&   -23.2&  -180.8&    25.4&$  U(-500,+500)$   \\ 
$\Delta V(CORAVEL-HIRES)$&[\ms]          &  4727&  4822&  4823&   124&  4682&  4965&  4542&  5106&$U(-4500,+5500)$\\ 
$\Delta V(C98-HIRES)$&[\ms]          &  4974.79&  4974.93&  4974.78&     1.76&  4973.04&  4977.01&  4971.25&  4979.41&$ U(-4500,+5500)$\\ 
$\Delta V(C07-HIRES)$&[\ms]          &  4973.50&  4972.78&  4972.91&     1.44&  4971.06&  4974.32&  4969.11&  4975.71&$ U(-4500,+5500)$   \\ 
$\Delta V(C14-HIRES)$&[\ms]          &  4989.71&  4987.44&  4987.91&     2.28&  4984.71&  4989.89&  4981.82&  4992.11&$ U(-4500,+5500) $ \\ 
\hline  
\multicolumn{9}{c}{ \bf Nuisance parameters}  \\ 
\hline  
$s(CORAVEL)$   &[\ms]          &    84&   176&    16&   105&    55&   314&     8&   385&$U(-500,+500)$ \\ 
$s(C98)$       &[\ms]          &     5.58&     6.10&     6.05&     0.68&     5.36&     6.90&     4.69&     7.79&$U(-30,+30)$   \\ 
$s(HIRES)$     &[\ms]          &     2.65&     3.94&     3.49&     1.16&     2.83&     5.43&     2.00&     7.33&$U(-30,+30)$   \\ 
$s(C07)$       &[\ms]          &     6.84&     6.64&     6.50&     0.61&     5.97&     7.36&     5.37&     8.16&$U(-30,+30)$   \\ 
$s(C14)$       &[\ms]          &     3.09&     3.35&     3.33&     0.48&     2.81&     3.91&     2.30&     4.51&$U(-30,+30)$   \\ 
\hline  
\multicolumn{11}{c}{ \bf HD4113Ab}   \\
\hline  
$\log{(P)}$    &log10([days])  &     2.721476&     2.721468&     2.721469&     0.000010&     2.721456&     2.721480&     2.721445&     2.721492&$U(0,+10)$     \\ 
$\log{(K)}$    &log10([\ms])   &     1.9504&     1.9515&     1.9503&     0.0038&     1.9473&     1.9559&     1.9429&     1.9603&$U(-10,+10)$   \\ 
$e$            &               &     0.9007&     0.9001&     0.8999&     0.0016&     0.8983&     0.9019&     0.8965&     0.9037&$U(0,0.985)$   \\ 
$\omega$       &[deg]          &   -48.64&   -48.40&   -48.45&     0.72&   -49.23&   -47.58&   -50.05&   -46.79&$U(0,360)$     \\ 
$M_{0}$        &[deg]          &    97.602&    97.585&    97.575&     0.051&    97.527&    97.643&    97.471&    97.703&$U(0,360)$     \\ 
\hline  
$P$            &[days]         &   526.594&   526.584&   526.586&     0.013&   526.570&   526.599&   526.556&   526.613&-              \\ 
$K$            &[\ms]          &    89.21&    89.43&    89.16&     0.78&    88.57&    90.34&    87.68&    91.26&-              \\ 
$A_{r}$        &[AU]           &     1.33&     1.298&     1.298&     0.026&     1.268&     1.328&     1.238&     1.355&-              \\ 
$m.\sin{i}$    &[Mjup]         &     1.674&     1.602&     1.602&     0.066&     1.527&     1.678&     1.452&     1.750&-              \\ 
$T_{p}$        &[day]          & 55642.768& 55642.740& 55642.738&     0.076& 55642.655& 55642.827& 55642.571& 55642.916&-              \\ 
$\lambda_{0}$  &[deg]          &    48.97&    49.19&    49.10&     0.69&    48.40&    49.97&    47.61&    50.72&-              \\ 
\hline 
\multicolumn{11}{c}{ \bf HD4113C}   \\
\hline 
$\log{(P)}$    &log10([days])  &     4.587&     4.582&     4.539&     0.084&     4.501&     4.689&     4.440&     4.819&$U(0,+10)$     \\ 
$\log{(K)}$    &log10([m/s])   &     2.539&     2.589&     2.581&     0.038&     2.549&     2.634&     2.513&     2.685&$U(0,+10)$     \\ 
$\sqrt{e}\,\cos{\omega}$&               &    -0.564&    -0.573&    -0.580&     0.055&    -0.633&    -0.507&    -0.689&    -0.448&$U(-1,+1)$     \\ 
$\sqrt{e}\,\sin{\omega}$&               &    -0.25&    -0.21&    -0.21&     0.10&    -0.32&    -0.08&    -0.40&     0.038&$U(-1,+1)$     \\ 
$T_{\rm Rvmin}$&[days]         & 47614& 47179& 47256&   475& 46585& 47661& 45883& 48043&$U(45000,80000)$\\ 
$i$            &[deg]          &    91.3&    89.1&    88.7&    10.6&    77.1&   101.3&    65.2&   112.9&$U(3,177)$     \\ 
$\Omega$       &[deg]          &    40.5&    41.8&    42.8&     6.4&    34.6&    49.0&    27.3&    56.5&$U(0,360)$     \\ 
\hline 
$P$            &[yr]           &   105&   105&    92&    22&    87&   134&    75&   180&-              \\ 
$K$             &[m/s]          &   346&   388&   379&    34&   354&   431&   326&   484&-              \\ 
$A_{r}$        &[AU]           &    23.7&    23.0&    21.1&     3.1&    20.3&    27.1&    18.5&    33.2&-              \\ 
$M_{C}.\sin{i}$&[\mjup]        &    59.7&    64.4&    63.5&     4.2&    59.8&    69.4&    55.8&    74.8&-              \\ 
$M_{C}$        &[\mjup]        &    59.8&    65.8&    64.7&     4.2&    61.4&    70.7&    57.4&    76.8&$U(0,M_{A}/2)$      \\ 
$e$            &               &     0.380&     0.377&     0.364&     0.063&     0.315&     0.458&     0.260&     0.543&$U(0,0.985)$   \\ 
$\omega$       &[deg]          &   203.9&   200.3&   203.9&     9.9&   188.1&   210.6&   176.4&   219.5&-              \\ 
$M_{0}$        &[deg]          &    63&    69&    55&    18.&    50&    92&    36&   113&-              \\ 
$T_{0}$        &[days]         & 48695& 48127& 48370&   652& 47327& 48818& 46518& 49384&-              \\ 
$\lambda_{0}$  &[deg]          &   -52&   -47&   -43&    14&   -64&   -33&   -80&   -20&-              \\ 
\hline  
\end{tabular}
\end{table}
\end{landscape}

\end{appendix}

\end{document}